\documentclass[12pt,preprint]{aastex}
\usepackage{lscape}
\usepackage{ulem}
\usepackage[usenames]{color}
\usepackage{graphicx}
\slugcomment{~~\today}
\newcommand       \msun        	{$M_{\odot}$}
\newcommand       \msune        	{M_{\odot}}
\newcommand       \lsun      	{$L_{\odot}$} 
 
\newcommand	     \mpc              {Mpc$^{-3}$}
\newcommand	     \cc             {cm$^{-3}$}

\newcommand	     \myr              {$M_{\odot}$~yr$^{-1}$}

\newcommand       \mic        	 {$\mu$m}

\newcommand      \jay     {J1148+5251}
\newcommand      \mav      {$\left<m\right>$}
\newcommand      \mave      {\left<m\right>}
\newcommand 	     \mstar      {$m_{\star}$}
\newcommand 	     \mstare      {m_{\star}}
\newcommand      \md     {$\left<m_d\right>$}
\newcommand      \mde     {\left<m_d\right>}

\newcommand 	     \taude      {\tau_d}

\newcommand      \rsne      {R_{\rm SN}}
\newcommand     \spitz    {{\it Spitzer}}
\newcommand	   \cals	{${\cal S}$}

\begin{document}

\title{THE ORIGIN OF DUST IN THE EARLY UNIVERSE: PROBING THE STAR FORMATION HISTORY OF GALAXIES BY THEIR DUST CONTENT}

\author{Eli Dwek\altaffilmark{1} and Isabelle Cherchneff\altaffilmark{2}}
\altaffiltext{1}{Observational Cosmology Lab, Code 665, NASA Goddard Space Flight Center,
Greenbelt, MD 20771; eli.dwek@nasa.gov}
\altaffiltext{2}{Department Physik, Universit\"at Basel, CH-4056 Basel, Switzerland; isabelle.cherchneff@unibas.ch}

\begin{abstract}
Two distinct scenarios for the origin of the $\sim 4\times10^8$~\msun\ of dust observed in the high-redshift ($z = 6.4$) quasar J1148+5251 have been proposed.  The first assumes that this galaxy is much younger than the age of the universe at that epoch so that only supernovae (SNe) could have produced this dust. The second scenario assumes a significantly older galactic age, so that the dust could have formed in lower-mass asymptotic giant branch (AGB) stars. Presenting new integral solutions for the chemical evolution of metals and dust in galaxies, we offer a critical evaluation of these two scenarios, and observational consequences that can discriminate between the two. We show that AGB stars can produce the inferred mass of dust in this object, however, the final mass of surviving dust depends on the galaxy's star formation history (SFH). In general supernovae cannot produce the observed amount of dust unless the average SN event creates over $\sim 2$~\msun\ of dust in its ejecta. However, special SFHs can be constructed in which SNe  can produce the inferred dust mass with a reasonable average dust yield of $\sim 0.15$~\msun. The two scenarios propose different origins for the galaxy's spectral energy distribution, different star formation efficiencies and stellar masses, and consequently different comoving number densities of \jay-type hyperluminous infrared (IR) objects.
The detection of diagnostic mid-IR fine structure lines, and more complete surveys determining the comoving number density of these objects can discriminate between the two scenarios. 

\end{abstract}

%================================================================ 
\section{INTRODUCTION}
%================================================================ 
Determining the origin of the massive amount of dust present in  SDSS~J1148+5251 (hereafter \jay), a hyper-luminous quasar at $z = 6.4$ presents a special challenge. The dust mass  inferred from far-infrared and submillimeter observations is $\sim (1-5)\times 10^8$~\msun\  depending on dust composition (Dwek et al. 2007, and references therein), about ten times larger than the mass of dust in the Milky Way \citep{sodroski97}. The total far-infrared (IR) luminosity is about $2\times 10^{13}$~\lsun, giving a star formation rate (SFR) of $\sim 3400$~\myr\, for a Salpeter initial mass function (IMF) \citep{kennicutt98a}. Since the universe was only $\sim 890$~Myr old at that redshift, and the galaxy perhaps much younger, it has been suggested that only core-collapse supernovae (CCSNe) can produce the observed amount of dust in this object \citep{morgan03,dunne03,nozawa03,maiolino04a,rho08,sibthorpe09}. Another argument made in favor of SNe as the most important dust sources in the early universe, was based on observations of unmixed ejecta of young remnants, primarily Cas~A. The inferred dust mass in this remnant, ranging from $\sim 0.1 - 0.15$~\msun, was considered sufficiently large for SNe to account for the observed dust mass in \jay\ \citep{rho08,sibthorpe09,barlow10}. We consider the $\sim 1$~\msun\ of dust mass claimed by \cite{dunne03, dunne09} to have formed in Cas~A as an unreasonable amount of dust. The total ejecta mass is $2-4$~\msun, of which only $\sim 0.2$~\msun\ consists of condensible elements [e.g. \cite{nozawa10}, and references therein].

None of these assertions, that SNe are significant dust sources in the early universe, were substantiated by detailed calculations. Specifically, the role of supernovae as destroyers of dust during the remnant phase of their evolution was completely ignored.  Furthermore, these claims tacitly assume that the epoch of intense star formation spanned the entire age of the galaxy, which requires an excessively large reservoir of interstellar gas. Finally, recent calculations show that only $\sim 0.15$~\msun\ of dust is created in a Population~III 20~\msun\ SN \citep{cherchneff10}, which may be a typical yield for Pop~II SNe of similar masses. All these considerations point to the need of more detailed calculations to ascertain the role of SNe as dust sources and sinks in the early universe. In such recent calculations, \citep{dwek07b} explored the combined effects of dust formation and destruction on the net amount of dust produced in \jay. They found that SNe cannot be important sources of dust in the early universe unless they produce significantly more, and destroy significantly less dust than implied by current observations or theoretical calculations.

In light of these difficulties,  \cite{valiante09} suggested that AGB stars could be the source of dust in \jay. Their model was motivated by the numerical simulations of the formation and growth of this galaxy through a series of successive mergers that resulted in repeated intense bursts of star formation [\citep{li07a}, hereafter Li07]. Associated with these mergers is the growth of its central black hole (BH) which at $z\approx 6.4$ has reached a mass of about $\sim 10^9$~\msun. In this scenario, star formation commenced at $z\approx 15$, when the universe was merely 250~Myr old. Consequently, the progenitors of the more numerous and efficient dust producing AGB stars had time to evolve off the main sequence (see Table 1), and produce the observed mass of dust in \jay.  

This paper takes a critical look at these two proposed dust formation scenarios, hereafter referred to as the SN and AGB scenarios. We start by introducing some basic definitions of the various quantities that govern the chemical evolution of galaxies, and present new integral solutions for the chemical evolution of their elemental and dust content (Section~2). In Section~3 we discuss the yields of the main stellar dust sources. These include the explosive SN ejecta, the fast winds created during the Wolf-Rayet (WR) stage of the evolution of stars with masses in excess of $\sim 40$~\msun, and the quiescent winds from AGB stars. In Section~4 we apply the model to follow the chemical evolution of \jay. We calculate the contribution of SNe and AGB stars to the production of dust in this galaxy, and the evolution of its stellar mass and luminosity. The simulated SFH of Li07 leading to the formation of \jay\ is not unique, and in Section~5 we examine alternate scenarios for the formation and survival of dust by AGB stars and SNe. Our results show that both, the AGB and the SN scenarios, are still viable for producing the mass of dust observed in \jay, each scenario with its own problems and limitations. We therefore present in Section~6 observational tests, including studies of the spectral energy distribution (SED), the rarity of \jay-type objects, and the inferred SFR, that can discriminate between the two.
In Section~7 we explore two additional, non-stellar, dust sources:  molecular clouds, which have always been considered as an environment for the growth and processing of interstellar dust grains in the Galaxy \citep{snow75,dwek80b, liffman89,dwek98,greenberg99,zhukovska08}, and AGN winds, which were recently suggested as potential producers of interstellar dust in quasars \citep{elvis02, maiolino06}. The results of the paper are briefly summarized in Section~8. 

Throughout this paper we adopt a flat $\Lambda$CDM cosmology, with a baryonic density parameter $\Omega_b=0.044$, a total matter (dark+baryonic) density parameter of $\Omega_m=0.27$, a vacuum energy density $\Omega_{\Lambda}=0.73$, and a Hubble constant of $H_0 = 70~$~km~s$^{-1}$~Mpc$^{-1}$ \citep{spergel07}.

%================================================================ 
\section{EQUATIONS FOR THE EVOLUTION OF DUST}
%================================================================ 
\subsection{Basic Definitions}
We define the stellar IMF, $\phi(m)$, so that $\phi(m) dm$, is the number of stars with masses between $m$ and $m+dm$, and normalize it to unity in the \{$m_l,\ m_u$\} mass interval,
 where $m_l$, and $m_u$ are, respectively, the lower and upper mass limits of the IMF. The IMF-averaged stellar mass, \mav, is then:
%--------------- eq 1
 \begin{equation}
\label{mav}
\mave = \int_{m_l}^{m_u}\ m\ \phi(m)\ dm \quad , 
\end{equation}
%--------------- 
The SFR, $\psi(t)$, is the mass of stars formed per unit time, and is related to the stellar birthrate, $B(t)$, by:
\begin{equation}
\label{birth_rate}
B(t)  = {\psi(t) \over \mave} .
\end{equation}
We assume that all stars in the [$m_l,\, m_w$] mass range end their life quiescently, whereas all stars with masses $m_w \leq m \leq m_u$ become CCSNe, where $m_w=8$~\msun, is their lower mass cut.
 
A useful quantity is \mstar, the mass of all stars born per SN event, given by:
\begin{equation}
\label{mstar}
\mstare \equiv \mave / \int_{m_w}^{m_u}\  \phi(m)\ dm
\end{equation}

\noindent
The SN rate, $R_{SN}$, is then given by:
\begin{equation}
\label{rsn}
\rsne(t)   =  B(t)\, \int_{m_w}^{m_u}\  \phi(m)\ dm = {\psi(t)\over \mstare} \qquad ,
\end{equation}

In all our calculations we will use a mass-heavy ISM characterized by a power law: $\phi(m) \sim m^{-\alpha}$ in the \{$m_l,\ m_u$\} = \{1~\msun,\ 100~\msun\} mass interval, with $\alpha=2.35$. For this IMF we get:
 \begin{equation}
\label{m}
\mave  =  3.1\ \msune  \qquad {\rm and} \qquad \mstare  =  53.0 \ \msune
\end{equation}

\noindent
The choice of this mass-heavy IMF is motivated by studies suggesting that star formation in the early universe was biased towards more massive stars \citep{bromm04}. For comparison, a Salpeter IMF gives values of $\mave \approx 0.35$~\msun, and $\mstare \approx 135$~\msun.
%----------------------------------------------------------
\subsection{Equations for the Evolution of Dust or Elements}
%----------------------------------------------------------
We first describe the equations for the evolution of the mass of dust in the ISM. Let $M_g(t)$ be the mass of the gas in the ISM at a given time $t$. The evolution of $M_d(t)$, the mass of dust in the ISM, is governed by the equation:
\begin{equation}
\label{dmddt1}
{dM_d(t)\over dt} = - Z_d(t)\, \psi(t) -{M_d(t)\over \tau_d(t)} +  {\cal S}(t)
 \pm \left[{dM_d(t)\over dt}\right]_{inf/out}
\end{equation}
where $Z_d(t)$ is the dust-to-gas mass ratio:
\begin{equation}
\label{zd}
Z_d(t) \equiv {M_d(t)\over M_g(t)} \quad.
\end{equation} 
The first term in Equation (\ref{dmddt1}) represents the rate at which the dust is removed from the ISM by star formation; the second is the rate at which dust is destroyed by sputtering or grain-grain collisions in SN blast waves. The parameter $\tau_d(t)$ is the timescale for the combined effect of these processes; the third term, \cals$(t)$, is a source function representing the rate of dust formation in the different astrophysical environments; and the fourth term represents the rate of increase/decrease in the dust mass as a result of infall/outflow from the galaxy. 

The timescale for grain destruction in supernova shock waves is given by \citep{dwek80b}:
\begin{equation}
\label{taud}
\taude \equiv {M_d \over \mde\, \rsne} 
\end{equation}
The parameter \md\ is the total mass of the refractory elements, initially locked up in dust, which are returned back to the gas phase of the ISM throughout the evolution of a single supernova remnant (SNR).  For the MW galaxy, \cite{jones04} found dust lifetimes of 400 and 600~Myr for silicate and carbon grains, respectively. We adopt here an average lifetime of $\tau_d\approx 500$~Myr. This average should be time dependent since the dust composition, specifically the relative abundance of silicate and carbon dust, evolves with time. However, such distinction is not warranted in the present investigation because of our current lack of knowledge of the nature of the dust in high-redshift galaxies.  Adopting a total dust mass $M_d \approx 3\times10^7$~\msun, and a Galactic SN rate $\rsne \approx 0.02$ \citep{diehl06}, gives a value of about 3~\msun\ for the grain destruction efficiency in the Galaxy. Since \md\ scales linearly with the mass of ISM dust, we can normalize it to its Galactic value:
\begin{equation}
\label{md}
\mde = \left[{Z_d(t)\over Z_{d,mw}}\right]\, \mde_{mw}
\end{equation}
where $\mde_{mw}=3$~\msun\ is the Milky~Way value of \md, and $Z_{d,mw} \approx 0.007$ is the average dust-to-gas mass ratio in the Milky Way \citep{zubko04}. \\
The grain destruction rate can then be written as:
\begin{equation}
\label{mdtau}
{M_d(t)\over \tau_d(t)}  =   Z_d(t)\,  \left[ {\mde_{mw} \rsne \over Z_{d,mw}}\right]
\end{equation}

The source function \cals(t) represents the rate at which the dust mass in the ISM increases by nucleation in the different dust sources, or by growth in the ISM, which is an important process responsible for the differential depletions in its different phases.  
Considering only stellar sources, the source function can be written as:

\begin{eqnarray}
\label{st}
{\cal S}(t) & = & \int_{\widetilde m(t)}^{m_w}\ \left[{\psi(t-\tau_{ms}(m))\over \mave}\right]\ Y_{d,agn}(m)\, \phi(m)\, dm  \qquad{\rm (AGB\ stars)}\\ \nonumber
& &  + \int_{m_{wr}}^{m_u}\ \left[{\psi(t-\tau_{ms}(m))\over \mave}\right]\ Y_{d,wr}(m)\, \phi(m)\, dm \qquad{\rm (WR\ stars)}\\ \nonumber
& &  + \int_{m_w}^{m_u}\ \left[{\psi(t-\tau_{ms}(m))\over \mave}\right]\ Y_{d,sn}(m)\, \phi(m)\, dm  \qquad{\rm (SNe}) \\ \nonumber
& &  + \, {\cal A}_{Ia}\int_0^t\ \left[{\psi(t-\tau)\over \mave}\right]\ Y_{d,Ia}\ f_{Ia}(\tau)\, d\tau \ \qquad \qquad{\rm (SNIa)}
\end{eqnarray}
\noindent
where the different terms represent the net contribution of the different stellar sources to the dust mass in the ISM. 
For each stellar mass, the SFR, $\psi(t')$, is calculated at the epoch $t'= t-\tau_{ms}(m)$, where  $\tau_{ms}(m)$ is the main sequence lifetime of a star of mass $m$ (Table 1). For AGB stars, the lower limit of the integral, ${\widetilde m(t)}$, is given by:  max\{$m_l, \, m'(t)$\}, where $m'(t)$ is the mass of a star with a main sequence lifetime $t$. The lower limit, $m_{wr}$ is the limiting mass above which stars undergo extensive mass loss and become WR stars.  $Y_{d,agb}(m)$, $Y_{d,wr}(m)$, $Y_{d,sn}(m)$, and $Y_{d,Ia}$ are the dust yields in the quiescent winds of AGB and WR stars, and in the explosive ejecta of core-collapse and Type Ia SNe, respectively. In the last term, ${\cal A}_{Ia}$ is a normalization constant that can be determined from the ratio between the observed frequency of Type~II and Type~Ia SNe, and $f_{Ia}(\tau)$ is the distribution function of the delay time, $\tau$, between the birth of the stellar system and the SNIa event \citep{greggio05, greggio10}. 

Finally, the  yields in eq. (\ref{st}), except the SNIa yields, depend on the initial stellar  metallicity. This implicit time dependence is suppressed in the equations, and its implementation in the solution is described in Section 2.4 below.

For stellar masses above $\sim 8$~\msun, $\tau_{ms}(m)<< t$ for most times of interest, so that the yield from SNe and WR stars can be set equal to $R_{SN}\, {\widetilde Y}_{d,sn}$, and $R_{WR}\, {\widetilde Y}_{d,wr}$, respectively. $R_{WR}$ is the rate of WR stars, which can be defined similarly to the SN rate [see eq. (\ref{rsn})] with $m_w$ replaced by $m_{wr}$. The IMF-averaged yields of SNe and WR stars are given by:
\begin{equation}
\label{yield}
{\widetilde Y}_{d,X} = \int_{m_X}^{m_u}\ Y_{d,X}(m)\, \phi(m)\, dm/\int_{m_X}^{m_u}\ \phi(m)\, dm
\end{equation}
\noindent
where $X\equiv sn$ or $wr$. 

The term describing the rate of change in the dust mass caused by the exchange of gas with the intergalactic medium depends on whether the gas is flowing in or out of the galaxy. In an outflow, the dust-to-gas mass ratio in the outflowing gas is about equal to the galaxy's value, $Z_d(t)$, and the term is given by: 
\begin{equation}
\label{out}
\left[{dM_d(t)\over dt}\right]_{out} = Z_d(t)\, \left({dM_g(t)\over dt}\right)_{out} \qquad{\rm for\ outflows}
\end{equation} 
In the case of infall, the dust-to-gas mass ratio of the infalling gas, $Z_d^{inf}(t)$, is an independent variable and the term becomes:
\begin{equation}
\label{inf}
\left[{dM_d(t)\over dt}\right]_{inf} = Z_d^{inf}(t)\, \left({dM_g(t)\over dt}\right)_{inf} \qquad{\rm for\ infalls}
\end{equation}

%----------------------------------------
\subsection{The Evolution of the Gas}
%----------------------------------------
The evolution of the ISM gas is given by the equation:
\begin{eqnarray}
\label{mgas}
{dM_g(t)\over dt} & = & - \psi(t) + \int_{m_l}^{m_u}\ \left[{\psi(t-\tau_{ms}(m))\over \mave}\right]\ M_{ej}(m)\, \phi(m)\, dm  \pm \left[{dM_g(t)\over dt}\right]_{inf/out} \\ \nonumber
 & \approx & -(1-R_{ej})\, \psi(t) \pm \left[{dM_g(t)\over dt}\right]_{inf/out}
\end{eqnarray}
where the first term in the top line represents the conversion rate of the ISM mass into stars, the second term represents the rate of stellar mass loss, where $M_{ej}(m)$ is the total mass returned back to the ISM by a star of initial mass $m$, and the third term represents the rate of change in the ISM mass as a result of gas infall or outflow. The second line represents the behavior of $dM_g(t)/dt$ in the instantaneous recycling approximation, where the returned (ejected) mass fraction, $R_{ej}$, is approximately 0.7.

The evolution of the gas mass is only weakly coupled to its chemical evolution. The returned mass, $M_{ej}(t)$, is almost independent of metallicity \citep{karakas07} and the stellar MS lifetimes are only weakly dependent on metallicity for values above 0.001. So with reasonable accuracy, the evolution of the gas mass can (but does not need to) be solved independently of the evolution of the dust and metals in the ISM.    

\subsection{Integral Solutions}

When the evolution of the ISM gas is decoupled from its chemical evolution, eq. (\ref{dmddt1}) can be readily solved with the aid of an integration factor, to yield a convenient functional form for the evolution of the dust or any element in the ISM.\\
Changing variables \citep{audouze76}: 
\begin{equation}
\label{dmddt2}
{dM_d(t)\over dt}  = M_g(t)\, {dZ_d(t)\over dt} + Z_d(t)\,  {dM_g(t)\over dt} \quad,
\end{equation}
 eq.~(\ref{dmddt1}) can be rewritten as:
%----------------
\begin{equation}
\label{dz_dt}
{dZ_d(t)\over dt} + Z_d(t)\, F(t) = G(t)
\end{equation}
%----------------
where $F(t)$ and $G(t)$ are dimensional functions (units of time$^{-1}$), given by:
%----------------
\begin{equation}
\label{Ft}
F(t) \equiv {\psi(t)\over M_g(t)}\, \left[\, 1+{\mde_{mw} \over \mstare\, Z_{d,mw}}+{1\over \psi(t)}\, {dM_g(t)\over dt} +  {1\over \psi(t)}\, \left({dM_g(t)\over dt}\right)_{out}\right]
\end{equation}
%----------------
and
%----------------
\begin{equation}
\label{Gt}
%G(t) \equiv {{\cal S}(t)\over M_g(t)}+{Z_d^{inf}\over M_g(t)}\, \left({dM_g(t)\over dt}\right)_{inf} \\
G(t) \equiv {1\over M_g(t)}\, \left[{\cal S}(t)+Z_d^{inf}(t)\, \left({dM_g(t)\over dt}\right)_{inf}\, \right]
\end{equation}
%----------------

\noindent
Note that the outflow term is included in $F(t)$ because it is proportional to $Z_d(t)$, whereas the infall term is included in the expression for $G(t)$, since it is independent of the evolving metallicity of the galaxy.
\noindent
Equation~(\ref{dz_dt}) can be solved with the aid of an integration factor $\lambda(t)$, which is defined as:  
%----------------
\begin{equation}
\label{lambda}
\lambda(t) = \exp\left[\int F(t)dt\right] \qquad ,
\end{equation}
%----------------
%\begin{equation}
%\label{zdt}
%Z_d(t) = \exp[-F(t)]\ \int_0^t\ \exp[F(t')]\, G(t')\, dt + {\cal C}
%\end{equation}
giving,
%----------------
\begin{equation}
\label{zdt}
Z_d(t) = \lambda(t)^{-1}\, \int_0^t\ \lambda(t')\, G(t')\, dt' + {\cal C}
\end{equation}
%----------------
\noindent
where ${\cal C}$ is an integration constant. \\ Given $M_g(t)$, the mass of dust is simply given by: 
\begin{equation}
\label{ }
M_d(t)=Z_d(t)\, M_g(t)
\end{equation}

Finally, equation~(\ref{dmddt1}) can also be used to describe the evolution of the metallicity, $Z_A$, of any stable or unstable element $A$ by the formal substitution of $M_d$ with $M_A$, and by using the appropriate yields in eq. (\ref{st}). For stable elements the parameter \md$_{mw}$ in eq.~(\ref{Ft}) should be set to zero, and for an unstable element, the dust lifetime, $\tau_d$, in eq. (\ref{dmddt1}) must be substituted by its radioactive decay time.  

The solutions, as written here, are easily calculated for yields at a {\it fixed} metallicity, when the source function $S(t)$ has a simple time dependency. Taking the changes of $S(t)$ with metallicity into account can be achieved in a straightforward way. Yields are usually calculated for a finite grid of  metallicities \{$Z_0(j), j=1,N$\}. One can therefore calculate the times \{$t_j, j=1,N$\} when the solutions derived with stellar yields at the fixed metallicities, $M[Z_0(j),t]$, reach the corresponding metallicities \{$Z_0(j), j=1,N$\}. The final evolution of the mass, $M(t)$, of the elements or the dust can then be written as a sum:
\begin{equation}
M(t) = \sum_1^N\ w_j(t)\, M[Z_0(j),t]
\end{equation}
where \{$w_j(t), j=1,N$\} are weight functions chosen to ensure the continuity of $M(t)$ across the grid of times \{$t_j$\}.
%=====================================================
\section{STELLAR SOURCES OF INTERSTELLAR DUST}
%=====================================================
In the following we examine the yields and relative importance of the dust formation in SNe, WR, and AGB stars. In all our calculations we adopt the previously described mass-heavy IMF.  
%------------------------------------------------------
\subsection{Supernovae}
%------------------------------------------------------
Consider a constant SN rate, $R_{SN}$. The dust production rate is simply given by $S_0 = {\widetilde Y}_{d,sn} \times R_{SN}$. 
Without grain destruction and with a SN rate of 20~yr$^{-1}$, SN need to condense only 0.03~\msun\ of dust over a period of $\sim 500$~Myr in order to produce an observed dust mass of $\sim 3\times10^8$~\msun.  
When grain destruction by SN is taken into account, the amount of dust produced by SNe is given by: $M_d \approx {\widetilde Y}_{d,sn}\, R_{SN}\, \tau_d$, giving the trivial solution: ${\widetilde Y}_{d,sn} \approx \mde$,
that is, in a steady state SNe must produce a dust mass that is equal to the amount they destroy during their lifetime. 
However, the required value of ${\widetilde Y}_{d,sn}$ can be lower before the system reaches a steady state. For example, the first SNe expand in a dust-free medium and therefore are net producers of interstellar dust. Detailed chemical evolution and population synthesis models are therefore needed to fit the observational constraints imposed by the inferred stellar, gas, and dust masses, as well as the stellar and radiative output of the galaxy.

Simple analytical models have shown that even in the absence of grain destruction, SNe must produce at least $\sim 0.3$~\msun\ of dust to account for the observed dust-to-gas mass ratio in \jay\ (Dwek et al. 2007; Figure 8). The reason that this number is not lower and equal to the value of $\sim 0.03$~\msun\ derived in our simple estimate above, stems from the fact that in these models the SFR is proportional to $\Sigma_{gas}^{1.5}$, where $\Sigma_{gas}$ is the mass surface density of the gas \citep{kennicutt98b}. $\Sigma_{gas}$ evolves with time, so that the effective timespan over which star-formation takes place is therefore much shorter than the age of the galaxy. It is determined by either the buildup of the ISM in infall models, or by the depletion of the ISM in closed box models. For \jay\ the situation is exacerbated by the fact that the IR-inferred SFR is $\sim 3000$~\myr. This SFR can therefore only be sustained for about 20~Myr, if the reservoir of gas is comparable to the dynamical mass or the CO mass of the galaxy. 

Do SNe actually produce this much dust? The most detailed information on the amount of dust formed in SNe comes from IR observations of the young supernova remnant of Cas~A. \spitz\ observations of the remnant have revealed the presence of $\sim 0.02-0.04$~\msun\ of hot \citep{rho08} and $\sim 0.08$~\msun\ of cool  dust \citep{sibthorpe09, barlow10}. Furthermore, detailed calculations, using a chemical kinetic approach to follow the transformation of gas phase molecules into small clusters of dust precursors, show that only $\sim 0.1-0.15$~\msun\ of dust is created in the explosive ejecta  of a Population~III 20~\msun\ SN \citep{cherchneff10}. Such yield may be typical of SNe of similar masses, regardless of the initial metallicity of the progenitor star. This yield is still not sufficient to account for the mass of dust observed in \jay\ except when grain destruction is almost nonexistent in that galaxy. Such special scenario is discussed in Section 4.1.  

%------------------------------------  
\subsection{Wolf-Rayet Stars}
%------------------------------------
Of the stars that become CCSNe, only those with masses above $m_{wr} \approx 40$~\msun\ will become WR stars, a stage during they will experience extensive mass loss.  The formation of carbon dust becomes feasible during the latest phases of this mass loss phase, when the stellar surface becomes carbon enriched, and the star evolves into a carbon-rich WR star (WC star) \citep{crowther97}. For an IMF with a Salpeter slope of $\alpha = 2.35$, the fraction of CCSNe that become WR stars is:
\begin{equation}
\label{fwr}
f_{wr} \approx \left(m_{wr}/m_w\right)^{1-\alpha} \approx 0.11
\end{equation} 
So WR stars have to produce about 10 times more dust than SNe to account for the dust mass in \jay. 
There is observational evidence that most of WR stars are part of a binary system with an OB companion and that the dust formation locus is located in the region where the two winds collide as exemplified by the archetypical systems WR104 and WR140. For the Pinwheel Nebula WR~104, a dust mass loss of $\sim 8\times 10^{-7}$~\myr\ is derived by \cite{harries04}. 
A typical 60~\msun WR star is characterized by a WC phase lasting about $2\times 10^5$~yr \citep{prantzos86}.  We can thus approximate the amount of dust formed by WR stars as $\sim 0.2$~\msun\ over their lifetime. This number is of the same order of magnitude and not 10 times larger than the dust mass formed in the ejecta of a CCSN. Therefore, WR stars are certainly contributing to the dust budget of \jay\ but no more than 10\% of all dust makers. 
  
%------------------------------------------------------
\subsection{AGB Stars}
%------------------------------------------------------
In the Milky Way, a significant fraction of dust is produced in AGB stars \citep{dwek98, tielens98}. Dependent of their evolutionary stage on the AGB, red giants have either a oxygen-rich stellar photosphere characterized by a C/O ratio $<$ 1, or a carbon-rich photosphere resulting from thermal pulses and 3rd dredge-up and characterized by a C/O ratio $>$ 1. The former stars form silicate and metal oxide dust whereas the latter essentially form carbon dust and silicon carbides. At present, there exist no satisfactory models explaining the chemical formation of dust in AGBs and the impact of dust on accelerating the outflow through radiative pressure. Dynamical models often use the classical nucleation theory to describe dust synthesis and derive dust yields \citep{ferrarotti06}, an approach that has been proved to be inadequate when applied to circumstellar outflows \citep{donn85,cherchneff10}.
 
\begin{figure}[htbp]
  \begin{center}
\includegraphics[width=3.2in]{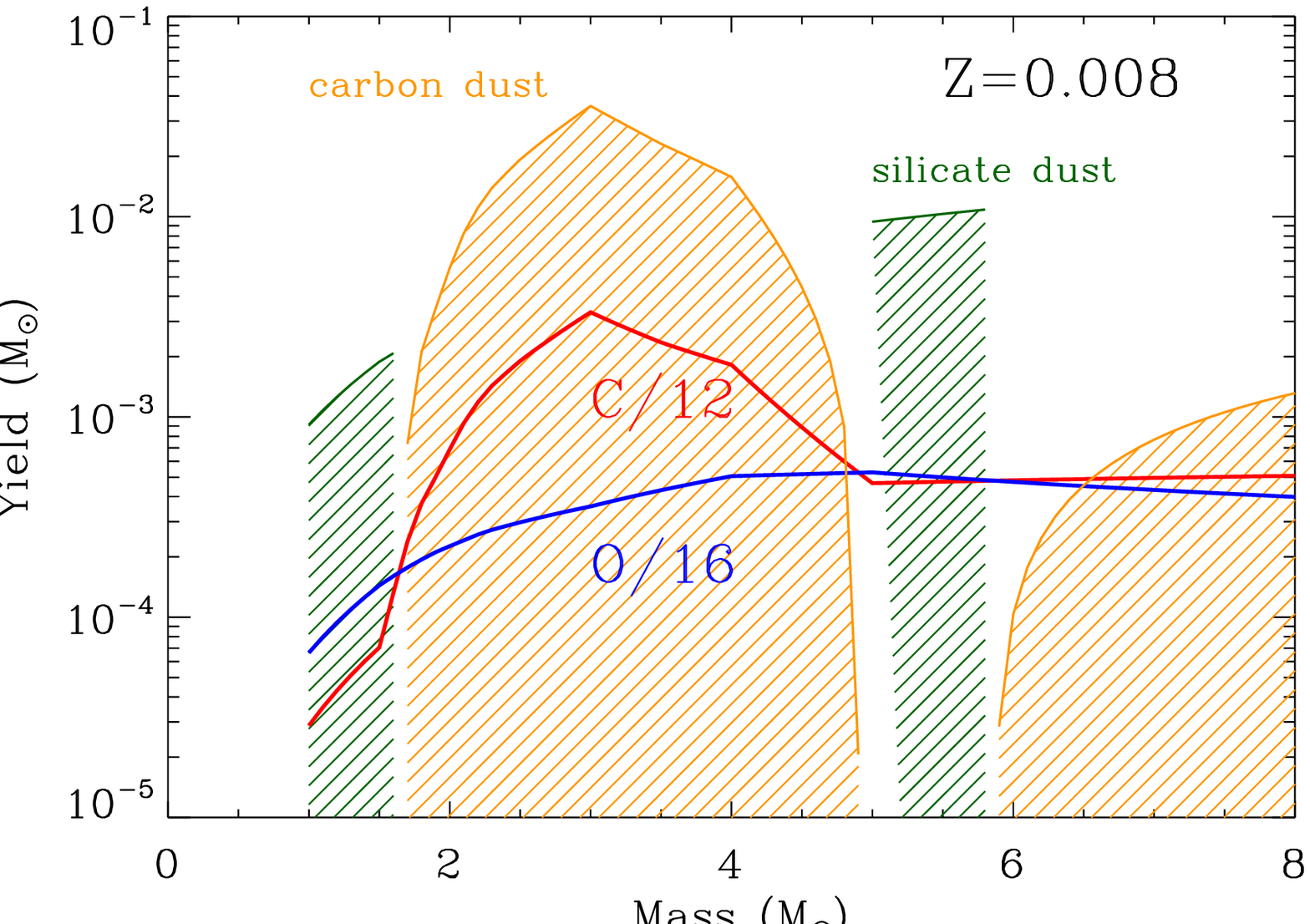}  
\includegraphics[width=3.2in]{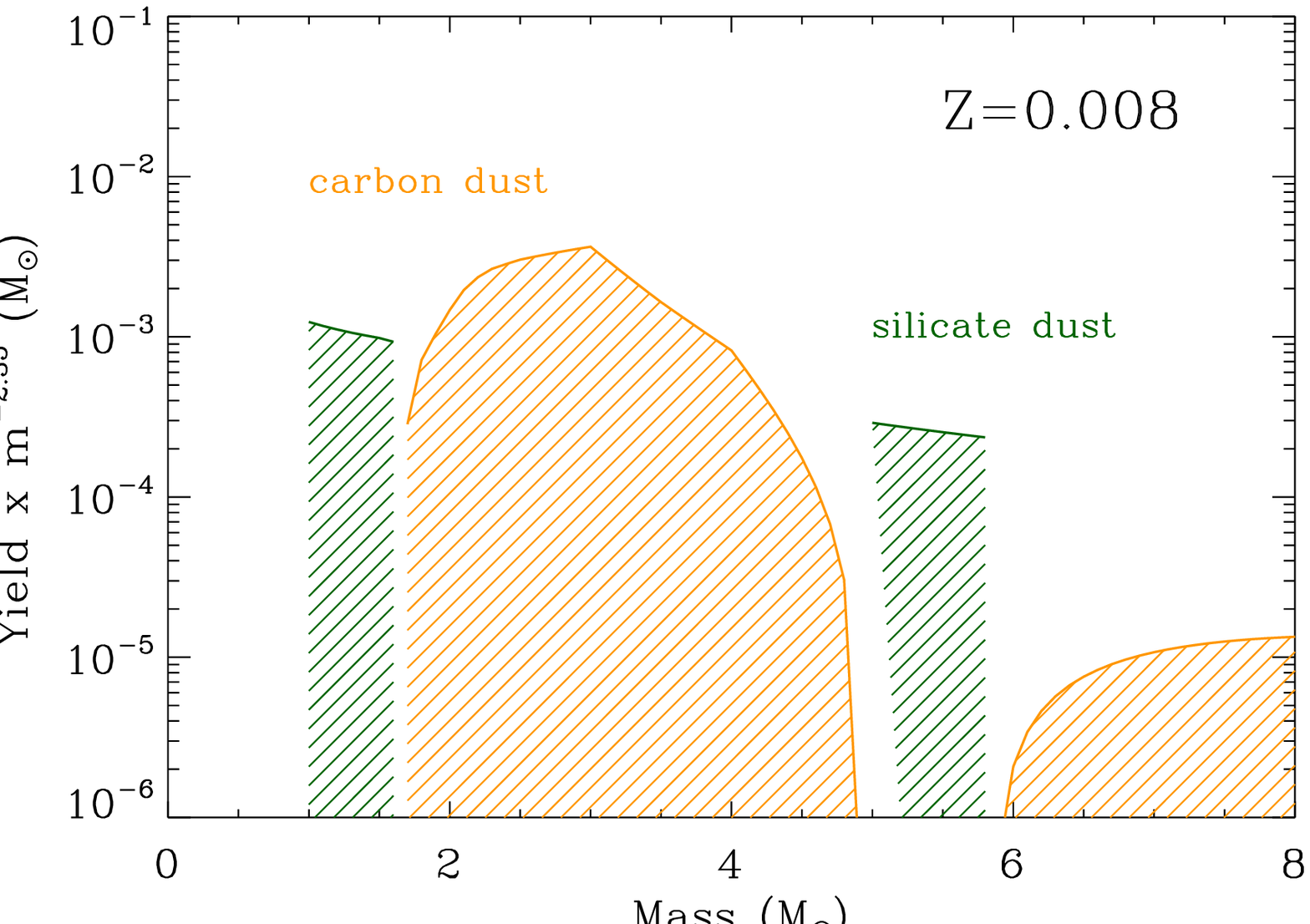}
\end{center}
 \caption{{\footnotesize {\bf Left panel}: The dust yield from AGB stars with an initial metallicity of 0.008 calculated using the yields of \cite{karakas07}. The red and blue lines depict the carbon and oxygen yields from these stars, divided by their atomic number. Carbon dust (hatched orange regions) is formed when the C/O ratio $>$ 1, whereas silicate dust (hatched green region) is formed when that ratio is $<$ 1. {\bf Right panel}: The IMF-weighted yield of dust from these stars. The $\sim 3$~\msun\ stars, which have a MS lifetime of $\sim 350$~Myr (see Table 1) are the dominant contributors to the production of dust. }}
    \label{agb_yield}
\end{figure}

For the purpose of the present study, we ignore the chemical and physical complexity of the dust formation processes and assume a condensation efficiency of unity to calculate the dust yields as described by \cite{dwek98}. The calculations assume that carbon dust forms when the C/O number ratio exceeds 1, and that silicate dust forms when this ratio is less than unity. This approach provides upper limits for the dust mass. For illustrative purposes, Figure \ref{agb_yield} (left panel) depicts the mass of dust produced during the AGB phase from 1-8~\msun\ stars with an initial metallicity of 0.008. Yields were taken from \cite{karakas07}. Also shown in the figure are the carbon and oxygen yields (in units of \msun\ and divided by their atomic mass), to illustrate the range of stellar masses that produce either silicate or carbon dust. Yields similar to those presented in this figure were adopted by \cite{valiante09} based on the models of \cite{ferrarotti06}. These models give different dust compositions, a distinction which is not important for the purpose of this paper.  

%------ figure 1

The right panel of the figure shows the IMF-weighted yield of dust from these stars, where the IMF is characterized by a power-law in mass with a Salpeter index of 2.35. The figure shows that he most efficient dust producers are $\sim 3$~\msun\ stars, which have a MS lifetime of about 350~Myr. For AGB stars to be significant contributors to the reservoir of dust in a galaxy, its age has to be significantly older than $\sim 400$~yr. 

%------------------------------------------------------
\subsection{Relative Importance}
%------------------------------------------------------
 The rate of dust production by SNe, WR stars, and AGB stars is given by: $\psi\times {\widetilde Y}_{d,sn} /\mstare$, $0.11\times\psi\times {\widetilde Y}_{d,sn} /\mstare$, and $\psi\times {\widetilde Y}_{d,agb} /\mave$, respectively. Taking 0.15, 0.2, and 0.04~\msun\ as the respective yields of dust in these sources, we get that in a steady-state, their relative dust production rate is: 1.0\ :\ 0.11\ :\ 4.6. AGB stars are therefore the most important sources of interstellar dust. However, this steady state is not immediately realized because of the delayed injection of AGB-condensed dust into the ISM. In the local group of galaxies, this delayed injection is manifested in the observed trend of the abundance of polycyclic aromatic hydrocarbons (PAHs) with galactic metallicity, which is taken as a proxy for galactic age \citep{dwek05,galliano08a}. The trend shows the existence of a threshold metallicity below which the abundance of PAH is very low. PAHs condense in the atmospheres of AGB stars, and therefore represent AGB-condensed dust. So at early times, the mass of dust in galaxies is dominated by SN-condensed dust. The dust evolution models presented in \citep{galliano08a} suggest that the steady state when AGB sources are the dominant dust producers may only be reached after about $\sim (1-2)$~Gyr of stellar evolution. PAHs can also be produced by shock processing of AGB-condensed carbon dust. However, both mechanisms, stellar condensation and shocks, need AGB stars as the source of dust, so the PAH-metallicity trend manifests the delayed injection of dust by AGB stars, even if most of the PAHs are formed in interstellar shocks. 
  
%====================================
\section{THE QUASAR J1148+5251}
%====================================

%------------------------------------------------------
\subsection{The Evolution the ISM Gas and Dust Masses}
%------------------------------------------------------
In the following we will apply our new mathematical formalism to calculate the evolution of the gas and dust in \jay. For AGB stars to be an important source of interstellar dust, star formation must have commenced at least 400~Myr before the epoch of observations which is about 900~Myr.  
In their paper, \cite{valiante09} adopted the SFH derived by Li07 from simulations of hierarchical galaxy mergers taking place at redshifts $z \gtrsim 6.5$.  The merger history of Li07 is not unique, and just one possible scenario leading to the formation of a supermassive BH at $z \approx 6$. We use it for illustrative purpose only, since it fulfills the necessary requirement for AGB stars to be significant sources of dust at high redshift. Their SFH, taken from their Figure 7 in Li07, is reproduced as a bold solid line in the left panel of Figure \ref{sfr}.

An important ingredient of chemical evolution models is the evolution of the gas mass. We adopted two different prescription for calculating this quantity. In the first, we adopted a closed box model (no infall or outflow) with an initial gas mass of $8\times10^{11}$~\msun, and used eq.~(\ref{mgas}) and the Li07 SFR to calculate its evolution. A value of $R_{ej}=0.48$ was chosen so that the stellar mass derived from the chemical evolution model would match that derived from the population synthesis model P\'EGASE (see Section 4.2). In the second approach we assumed that the SFR derived from the galaxy simulation follows the Schmidt-Kennicutt law in which $\psi(t) = A\, M_{gas}^{1.5}(t)$, where $A$ is a proportionality constant and a measure of the star formation efficiency in the galaxy. The value of $A$ was chosen to reproduce the model's SFR of 60~\myr\ at $t=900$~Myr with a gas mass of $\sim 5\times10^{10}$~\msun.   

In the calculations we adopted an IMF-averaged dust yield of ${\widetilde Y}_{d,sn} = 0.15$~\msun, the yield derived by \cite{cherchneff10} for a 20~\msun\ primordial core-collapse supernova. The yield of dust in AGB stars was calculated as in \cite{dwek98} using the stellar elemental yields of \cite{karakas07}. The total dust yield from AGB stars is only weakly dependent on metallicity, so for simplicity we adopted AGB yields at a constant metallicity of $Z=0.008$.  The stellar IMF was chosen to be a power law: $\phi(m) \sim m^{-\alpha}$ between 1 and 100~\msun, with a spectral index of $\alpha = 2.35$. 

%------ figure 2
\begin{figure}[htbp]
\vspace{0.2in}
  \centering
      \includegraphics[width=3.2in]{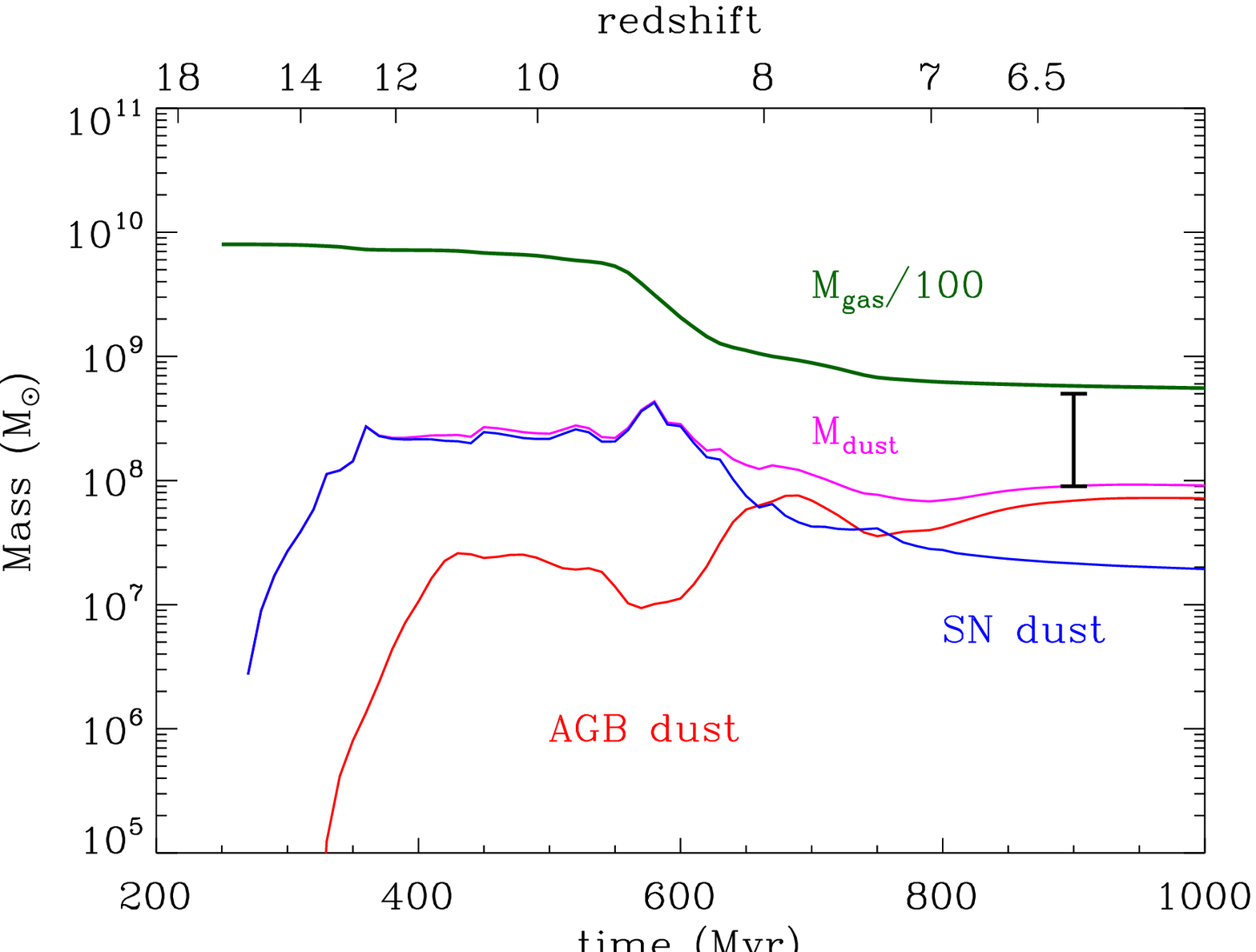}
  \includegraphics[width=3.2in]{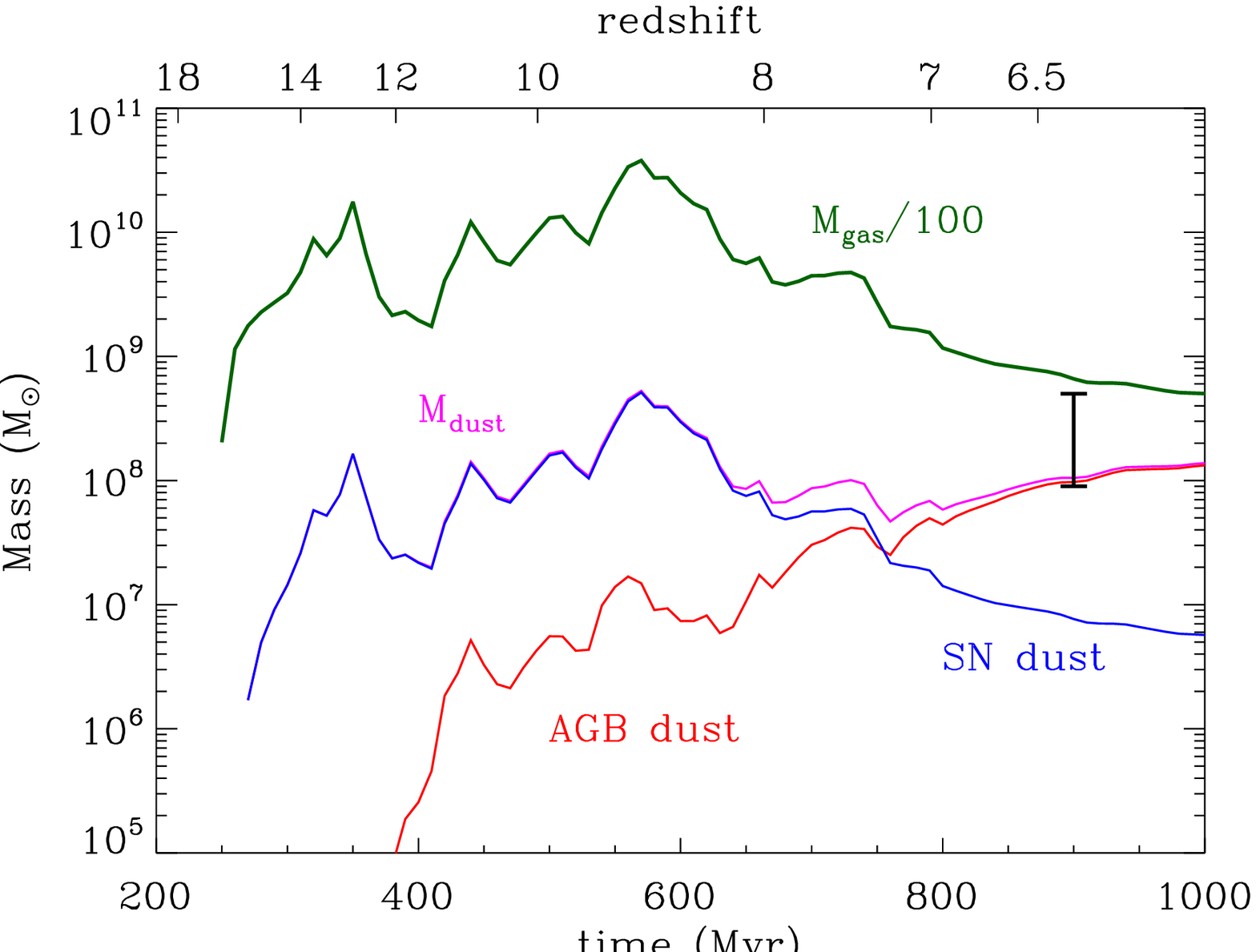}
 \caption{{\footnotesize The evolution of the mass of the SN-condensed dust, AGB dust, total dust mass, and the mass of the ISM gas in \jay, for the SF scenario of Li07, depicted here in Fig.~(\ref{sfr}). {\bf Left panel}: Results for the closed box model. {\bf Right panel}: Results for the Schmidt-Kennicutt SF law. The vertical line at $t=900$~Myr is the observationally-inferred dust mass.}}
\label{dustvol}
\end{figure}

Figure \ref{dustvol} depicts the evolution of dust and gas masses for the two adopted evolutionary models for $M_g(t)$: the closed box model (left panel), and the model described by the Schmidt-Kennicutt SF law (right panel). The vertical line at $t=900$~Myr is the observed dust abundance, the length of the bar reflecting the range of dust masses for the different dust compositions \citep{dwek07b}. The results of our calculations are in good agreement with those of \cite{valiante09}. Not surprising, the detailed evolution differs from theirs, since it depends on their adopted evolutionary history of the gas mass, the grain destruction efficiency, and AGB yields, all quantities that differ somewhat from the ones used in this work. However, the main results, the final masses produced by SN- and AGB-condensed dust, are very similar. The figures show that AGB stars can readily produce the observed amount of dust with the proposed SFR, and that SNe contribute less than 20\% of the total dust mass at $t = 900$~Myr. They can therefore only be important dust sources if their IMF-averaged yield is $\sim 1 $~\msun, confirming the results of the simpler analytical model of \cite{dwek07b}. 

Figure \ref{dustvol} also shows that for the Schmidt-Kennicutt model  there exist two epochs, occurring $\sim 100$ and $\sim 400$~Myr after the commencement of SF at $z\approx 15$, during which SNe produce over $\sim 10^8$~\msun\ of dust. It therefore seems that in a SF scenario consisting of either an extremely young or an intermediate-age burst of stars, SNe can produce the observed amount of dust, even with an average dust yield of $\sim 0.15$~\msun, without resorting to the need of AGB stars.

At the time of the intermediate burst, the SFR is over $10^4$~\myr. Compared to the younger burst, this burst requires a significantly larger SFR to produce a similar amount of dust, a consequence of the effect of grain destruction. The corresponding bolometric luminosity is over $2\times 10^{14}$~\lsun. Both, the SFR and the luminosity are in excess of the observational constraints, ruling out this SN scenario as a viable source of dust.  

\begin{figure}[htbp]
\vspace{0.2in}
  \centering
 \includegraphics[width=3.2in]{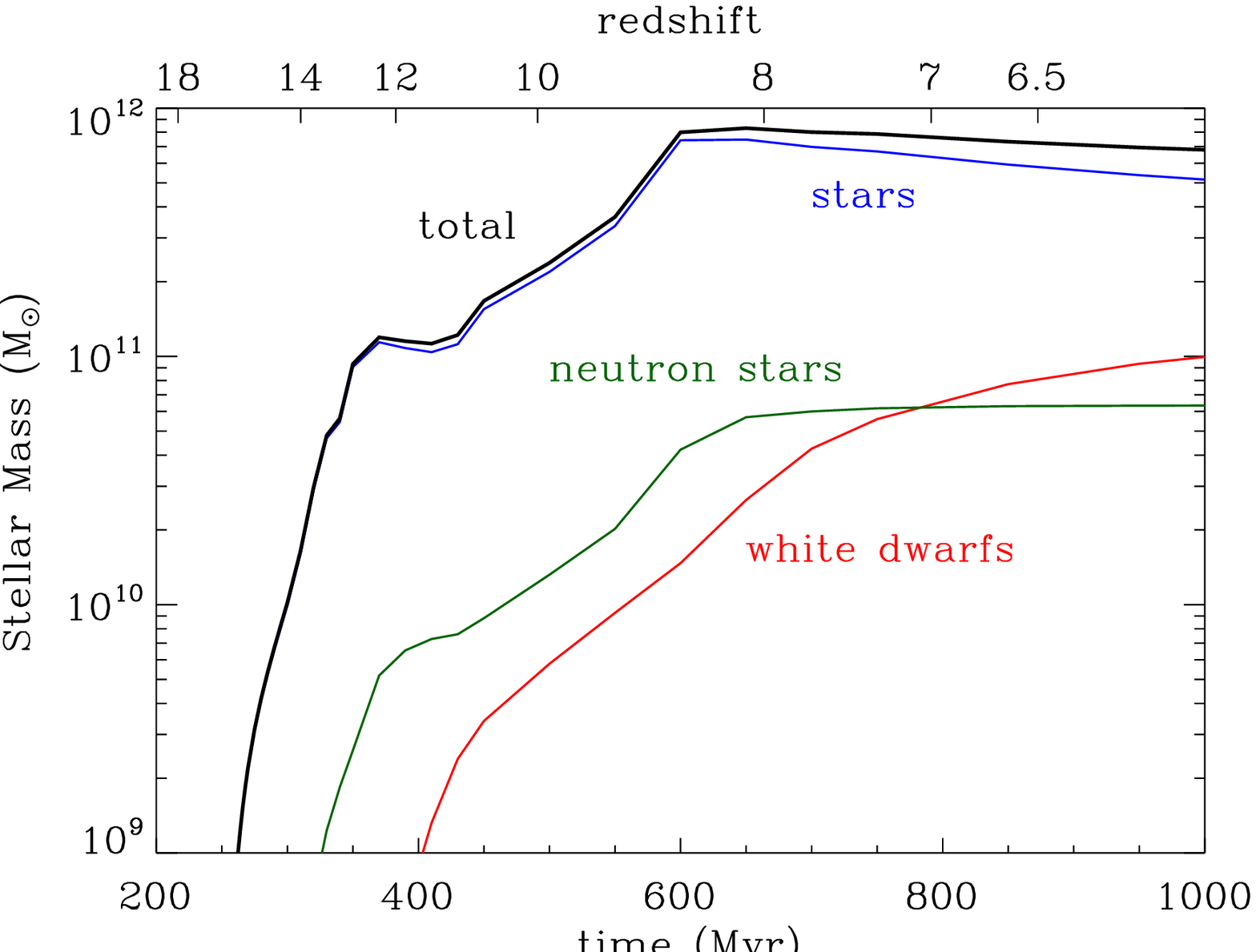}
    \includegraphics[width=3.2in]{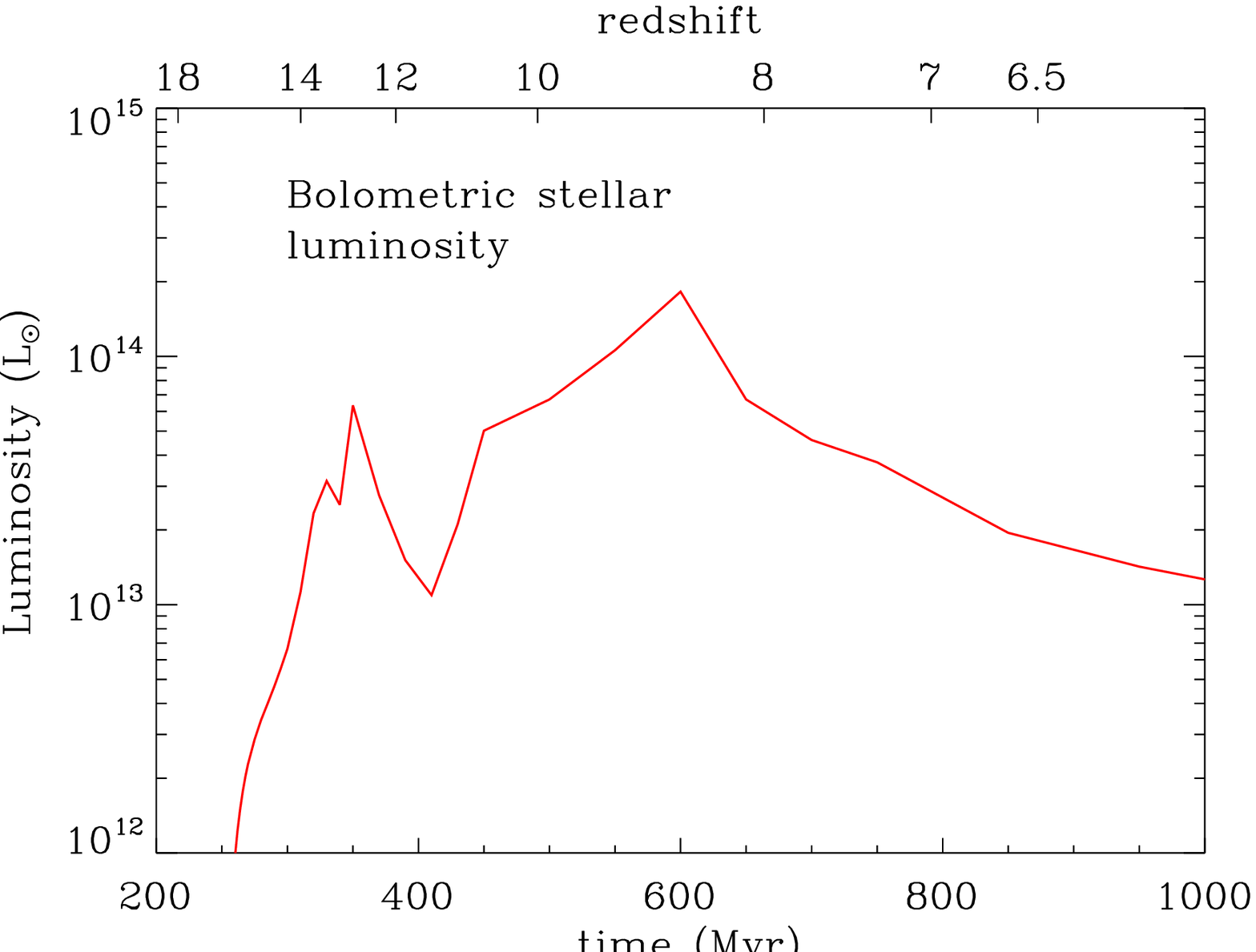}
  \caption{{\footnotesize {\bf Left panel}: The evolution of the stellar mass content (left panel) and luminosity (right panel) calculated with the population synthesis code P\'EGASE using the hierarchical SFH depicted in Figure~\ref{sfr}.}}
\label{starvol}
\end{figure}

The extremely young burst will produce a total bolometric luminosity of $\sim 5\times10^{13}$~\lsun, similar to the observed luminosity of $\sim 10^{14}$~\lsun, and a stellar mass of $\sim 10^{11}$~\msun\ (see Figure~\ref{starvol}). This is a viable scenario for the formation of dust in \jay, although it requires the formation of supermassive BH on a very short timescale, and has implication on the comoving number and luminosity densities of such objects in the early universe (see Section 6.2 below).

%-----------------------------------------------------------
\subsection{The Evolution the Stellar Luminosity and Mass}
%-----------------------------------------------------------
Figure \ref{mstar} depicts the evolution of the stellar masses and remnants (left panel) and the bolometric stellar luminosity (right panel) of \jay. Calculations were done for the hierarchical SF history depicted in figure~\ref{sfr}, using the P\'EGASE population synthesis code \citep{fioc97}. Similar figures with identical results were presented by Li07 (Figures 11-12 in their paper). The merger history was proposed by them to account for the presence of a supermassive black hole (BH) of mass $\sim 3\times 10^9$~\msun\ \citep{willott03}. Assuming that the relationship between the BH mass and its spheroidal component is still valid at redshifts $\gtrsim 6$, gives a stellar bulge mass of $\sim\, $few$\, \times10^{12}$~\msun\ [e.g. review by \citep{kormendy01}].  While the stellar mass derived here agrees with that estimate, it is in excess of the dynamical mass of $\approx (5.0\pm2.5)\times10^{10}$~\msun\ that is enclosed within a 2.5~kpc radius of \jay\ \citep{walter04}. \cite{walter04} proposed several solutions to this discrepancy, including an overestimate in the mass of the BH, and a breakdown in the assumption that the CO gas is gravitationally bound. Also, there may be evidence that at high redshifts BHs may grow very rapidly without the corresponding increase in the mass of the host galaxy \citep{shields06a}. It is therefore premature to rule out the AGB dust formation scenario while the origin of this discrepancy is still unresolved.  

%------ figure 3

%----------------------------------------------------------------------
\section{THE DEPENDENCE OF DUST MASS ON THE \\ STAR FORMATION HISTORY}
\begin{figure}[htbp]
  \centering
 \includegraphics[width=5in]{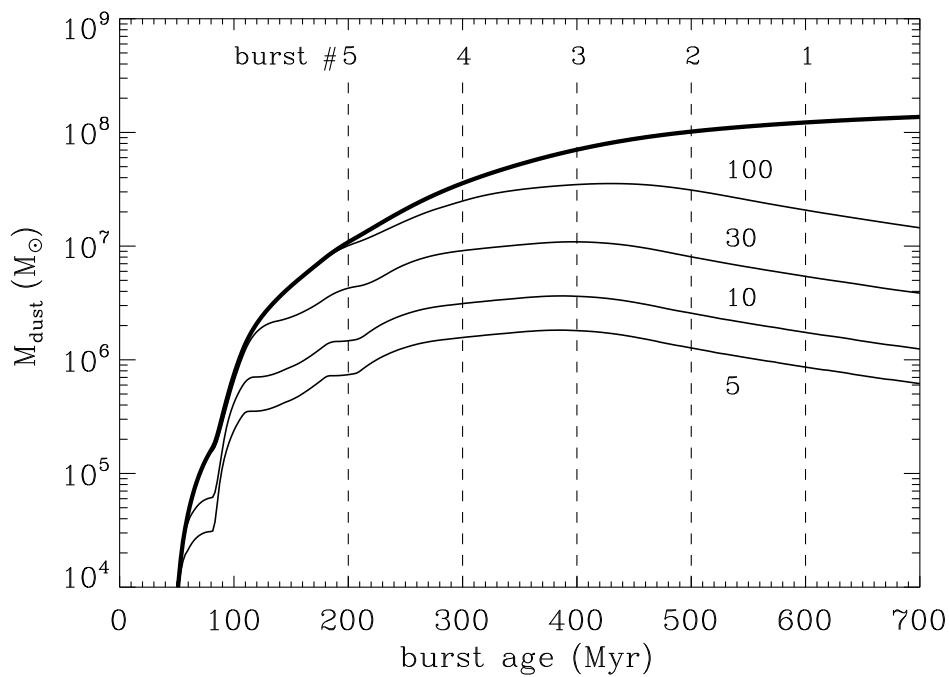}
  \caption{{\footnotesize  The evolution of the mass of dust released by AGB stars in a burst of star formation that commenced at $t=0$ as a function of time. The burst duration is 100~Myr, and its intensity is 1000~\myr. The top bold line represents the evolution of the dust mass without any grain destruction. Other lines depict the evolution of the dust when it is destroyed by a constant ongoing destruction mechanism, assumed to have started after the end of the burst. The lines are marked by the dust lifetime in units of Myr. The vertical dashed lines mark the age of the burst at epochs that correspond to the burst ages in Fig.~\ref{sfr}. }}
\label{burst}
\end{figure}

%----------------------------------------------------------------------
As pointed out earlier, the Li07 SF scenario for the formation of \jay\ is not unique. We therefore explore the dependence of the total dust mass on the SFH of the galaxy.
 We first consider the production and survival of dust by single short bursts of star formation. Figure~\ref{burst} depicts the evolution of the mass of dust produced by AGB stars that were born at time $t=0$ in a 100~Myr burst of star formation characterized by a constant SFR of 1000~\myr. The bold solid line depicts the net production of dust by AGB stars when no grain destruction occurs. It takes about $\sim 50$~Myr for the first AGB stars to evolve off the main sequence. The thinner lines depict the dust evolution in the presence of grain destruction. Grain destruction was assumed to be constant and to commence only after the cessation of the burst. 
After 600~Myr the burst releases $\sim 1\times10^8$~\msun\ of dust if no destruction has taken place in the intervening period since the end of the burst and the time of observation. The surviving mass of dust is significantly smaller when grain destruction is taken into account. A young burst produces after 200~Myr the least amount of dust, about $10^7$~\msun\ with no grain destruction, but is also least affected by grain destruction. The figure shows that a single burst of star formation (with no subsequent grain destruction), for example burst~1 with an intensity of $\sim 3000$~\myr, or burst 4 with an intensity of 10,000~Myr, can produce the observed $\sim 3\times10^8$~\msun\ of dust in \jay. 

%------ figure 4

\begin{figure}[htbp]
  \centering
 \includegraphics[width=3.2in]{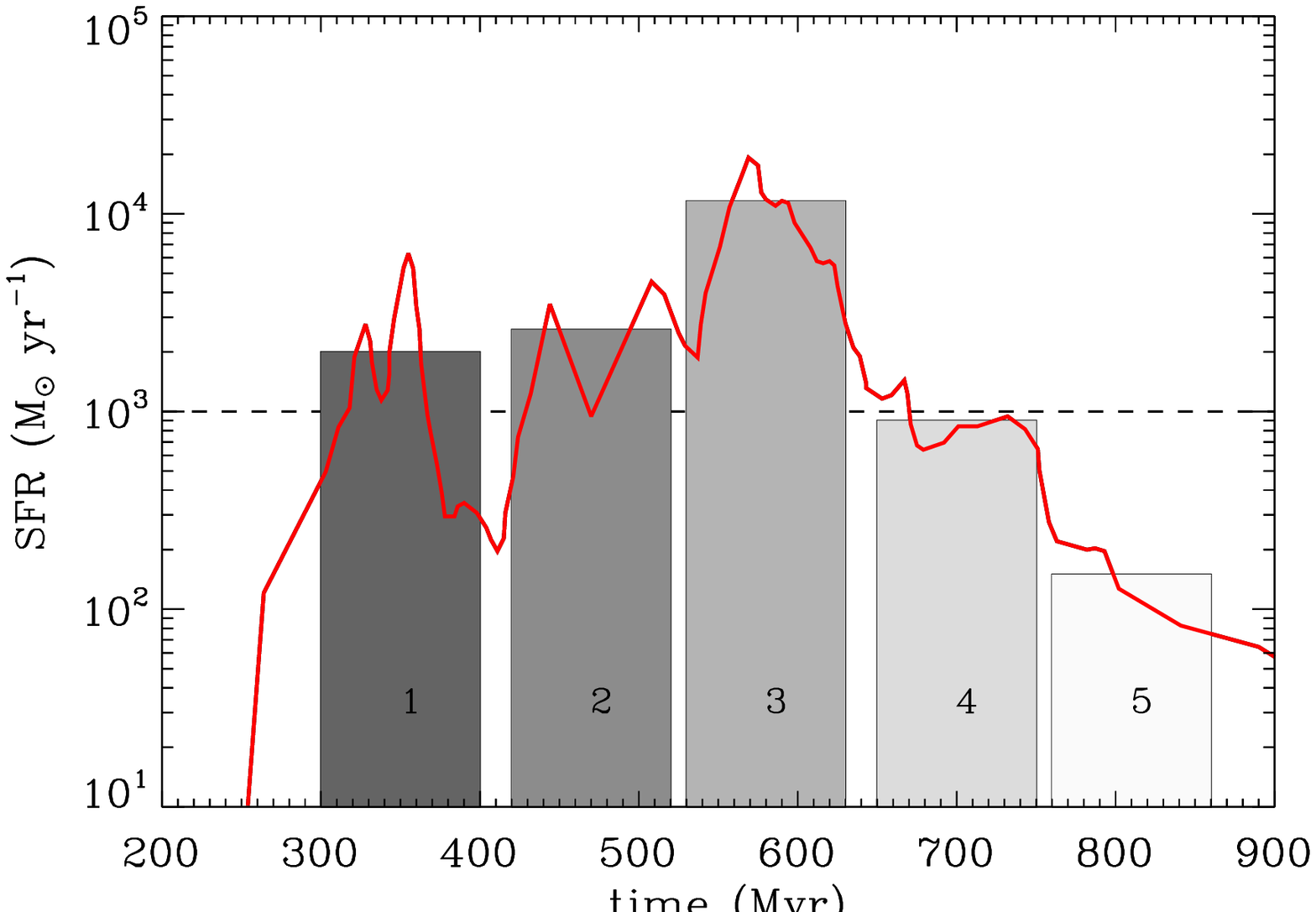}
    \includegraphics[width=3.2in]{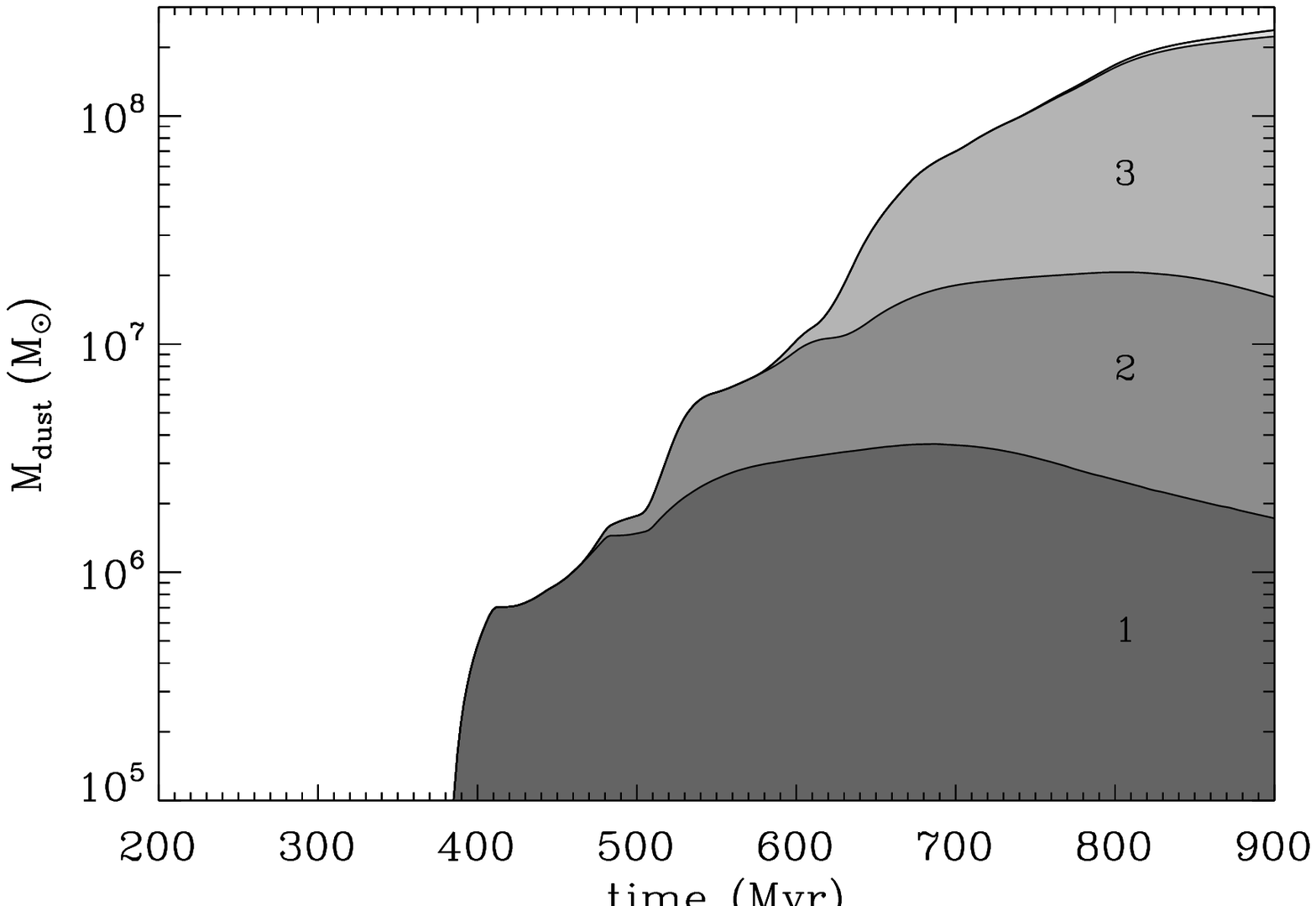}
  \caption{{\footnotesize {\bf Left panel}: The star formation rate of a galaxy going through hierarchical merger episodes (Li07, see also Figure 1 in \citep{li08}) shown as a bold red line, can be approximated by a series of 5 discrete 100~Myr duration bursts (shaded bars). {\bf Right panel}: The contribution of each of the five bursts depicted in the left panel to the total dust mass observed at the galactic age of 900~Myr. Most of the dust is that created by the third burst which had the largest star formation rate, and is followed by a period of relatively low rates of star formation and grain destruction. The figure illustrates the sensitivity of the surviving dust mass to the SFH of the galaxy.}}
\label{sfr}
\end{figure}

Deconstructing the complex SFH of a galaxy undergoing a series of hierarchical mergers can be very illustrative for understanding the origin of dust in such systems. In Figure~\ref{sfr} (left panel) we approximate the SFH of \jay\ used by \cite{valiante09} by five discrete bursts of star formation. The right panel depicts the cumulative contribution of each of the bursts to the total mass of dust in the galaxy. The oldest burst (number~1) releases the most amount of dust since all stars above 2.1~\msun\ contributed to its production. However, grain destruction by the subsequent bursts reduces the mass of dust that survives until 900~Myr  to only $\sim 2\times 10^6$~\msun. Almost all of the surviving dust was produced by the third burst. It has a main sequence turnoff mass of $\sim 3$~\msun, but also the largest SFR, and it is followed by a period of relatively low star formation activity and grain destruction.

The formation of about $\sim 3\times 10^8$~\msun\ of dust at $z \gtrsim 6$ can therefore be achieved for different SFHs. As illustrated in Figure~\ref{burst}, a single burst of star formation with the appropriate intensity can produce the required amount of dust, since no grain destruction takes place after the cessation of all star formation activities in the galaxy. Massive amounts of dust can also be produced by complex SFHs such as that presented in Figure~\ref{sfr}, provided that the epochs of intense starburst acivity are followed by a period with a very low rate, $\lesssim 70$\myr, of star formation, so that the dust lifetime will be over 100 Myr. 
 The discriminating factor between the different star formation histories capable of producing large amount of dust are the cumulative products, such as the metals or stellar masses, resulting from the star formation activities.

%------ figure 5

\newpage
%====================================
\section{OBSERVATIONAL DISCRIMINATION BETWEEN \\ THE SN AND AGB SCENARIOS}
%====================================
We have shown that not only AGB stars, but also SNe can produce the inferred dust mass in \jay, even with observed SN dust yields inferred from observations of Cas~A. The SN scenario requires a special SF event, characterized by an intense burst of star formation that commenced just prior to the observations. The short duration of the burst can insure that grain destruction did not have any time to significantly erode the newly-formed dust. This is clearly an idealized scenario, since some grain destruction will take place when the reverse shock travels through the SN ejecta \citep{nozawa10}. In the following we discuss several observational effects that can discriminate between the two scenarios. 

%------------------------------------------------------------
\subsection{The Spectral Energy Distribution}
\begin{figure}[htbp]
  \centering
  \includegraphics[width=3.2in]{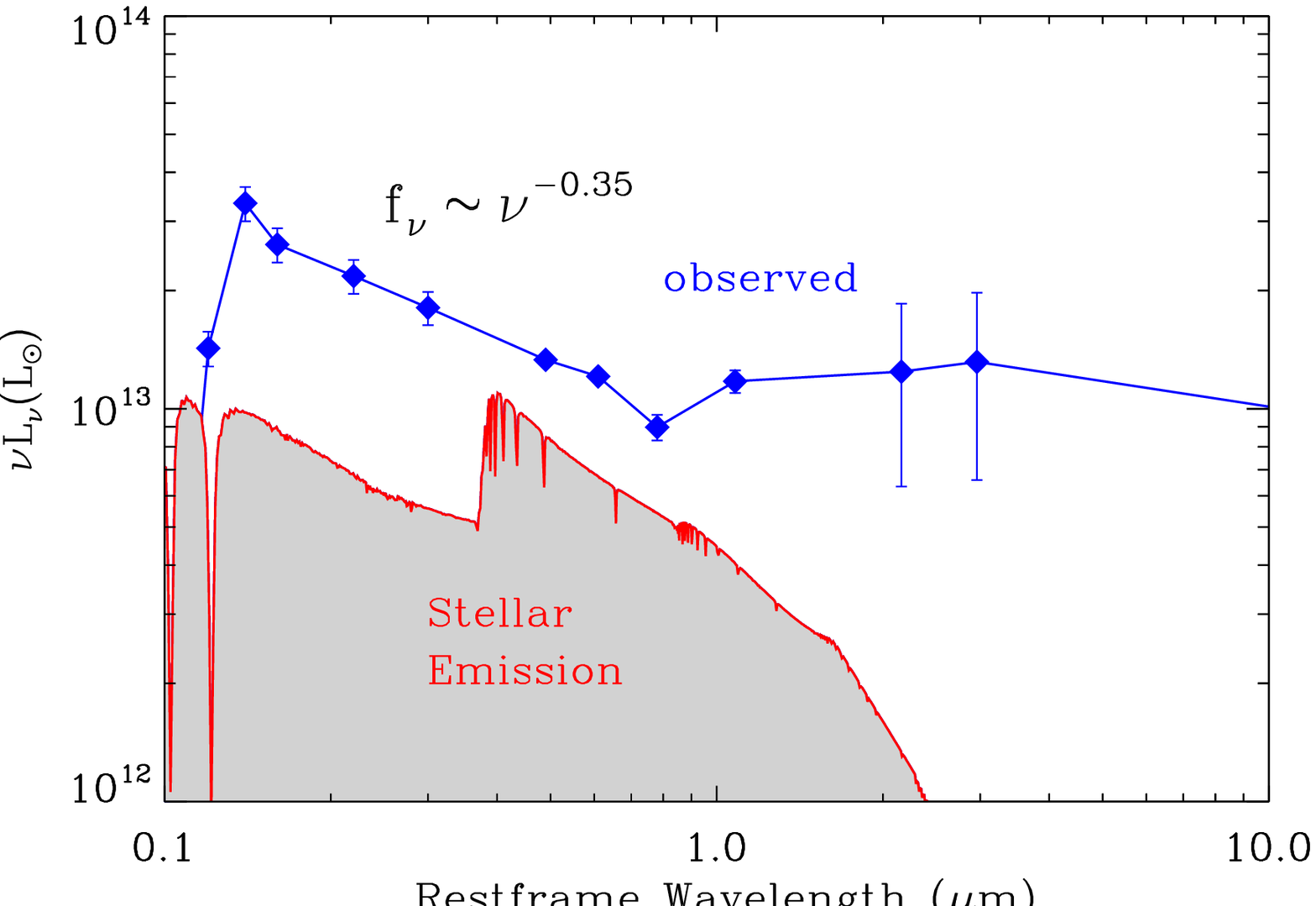}
    \includegraphics[width=3.2in]{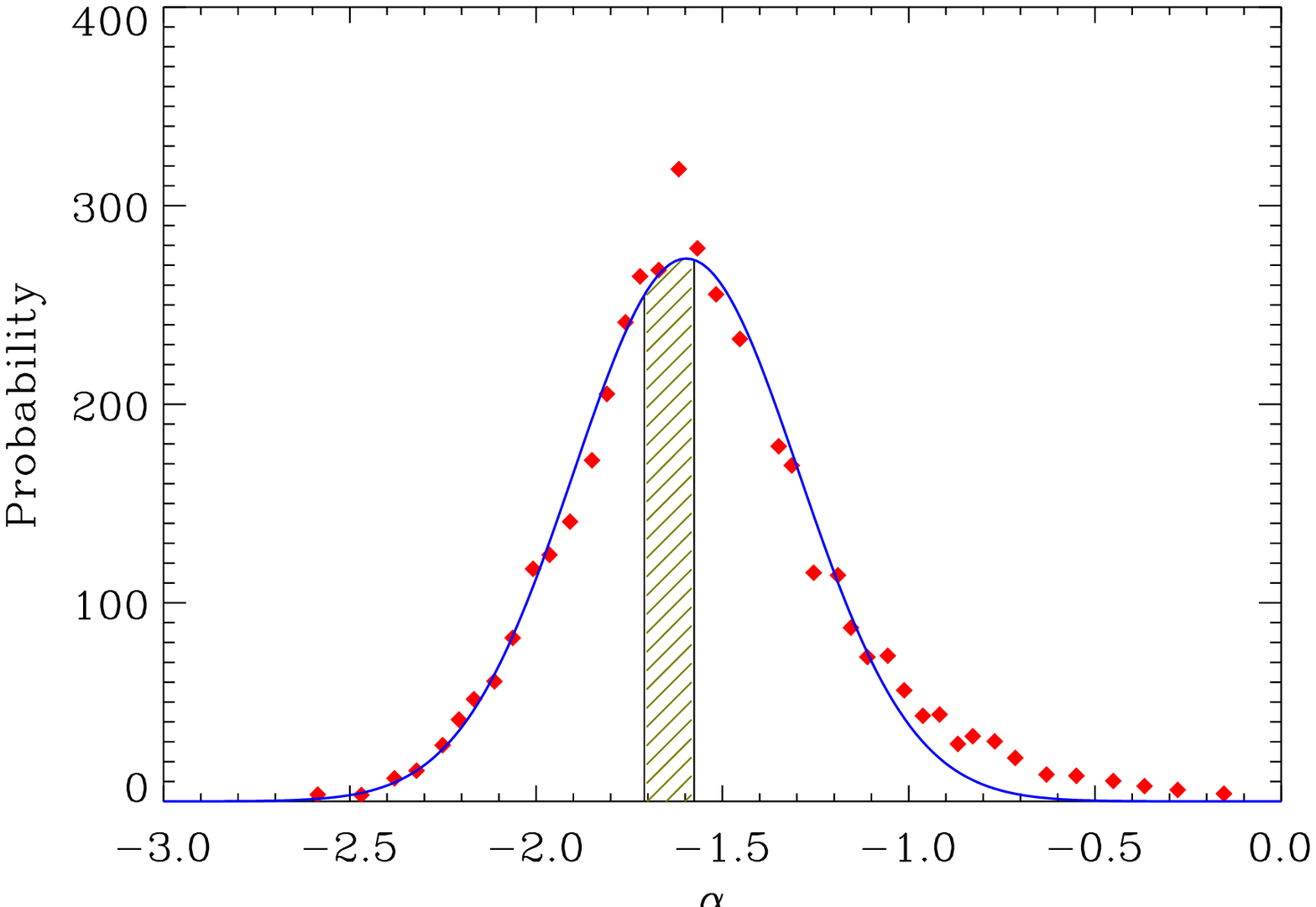}
  \caption{{\footnotesize {\bf Left panel}: Comparison of the stellar SED calculated for the hierarchical merger SFR at $z$=6.4 (shaded area) with the observations. In the AGB most of the optical to near-IR flux is emitted by the AGN. In contrast, in the SN scenario for the origin of the dust, most of this emission originates from stars [see Fig 11 in \citep{dwek07b}]. {\bf Right panel}: The probability distribution of photometric spectral indices $\alpha_p$ derived from the observed magnitudes of quasars \cite{richards03}. The spectral index of $\alpha_{\nu} \approx -0.35$ (left panel) corresponds to a value of $\alpha_p \approx -1.65$. The figure illustrates the fact that there is a high probability that the observed 0.1--1~\mic flux could be entirely attributed to the emission from an AGN. }}
\label{sfr_merger}
\end{figure}

%------------------------------------------------------------
An important distinction between the SN and AGB scenarios for the origin of dust in \jay\ is the origin of its SED. In the SN scenario, most of the UV-optical and all the far-IR emission originate from starlight and starlight-heated dust. In the AGB scenario the situation is more complicated, and the origin of the SED depends on the fractions of the AGN and starburst luminosities that are absorbed and re-emitted by the dust. The total calculated stellar luminosity at $z=6.4$ ($t\approx 900$~Myr) is $1.6\times10^{13}$~\lsun\ (see Figure \ref{starvol} and Table \ref{lum}), about equal to the total far-infrared (FIR) luminosity, which is $\sim (2.0\pm0.5)\times10^{13}$~\lsun. So, in principle, all the FIR emission could arise from dust heated by the stellar radiation. In this extreme case, {\it all} the observed UV to near-IR luminosity should originate from the AGN. Figure \ref{sfr_merger} (left panel) depicts the
stellar SED calculated for the hierarchical merger SFR at the galactic age of 900~Myr. In the absence of any discriminating line emission, can this extreme case be ruled out on the basis of the 0.1 to 1~\mic\ continuum emission? This question is addressed in the right panel of the figure which depicts the probability distribution of photometric spectral indices $\alpha_p$ derived from the observed magnitudes of quasars \cite{richards03}. The spectral index of the 0.1--1.0~\mic\ spectrum of \jay\ is $\alpha_{\nu} \approx -0.35$ (left panel) which corresponds to a value of $\alpha_p \approx -1.65$. This value is near the median value of -1.6. The optical to near-IR spectrum of \jay\ has therefore an equally high probability of being entirely that of an AGN as that of a starburst galaxy. The nature of the SED can therefore only be determined with the detection of mid-IR fine structure lines from highly ionized species such as [Ne V]14.3 and 24.3~\mic\ which are unambiguous diagnostics of AGN activity \citep{satyapal09,dudik09}. 
 
 %------ figure 6

%--------------------------------------------
\subsection{The Rarity of Dusty Hyperlumimous IR Galaxies at High Redshifts}
%--------------------------------------------
The AGB scenario for the formation of dust in \jay\ requires the early onset of star formation. \cite{valiante09} used the SFR derived in the Li07 hierarchical mergers model as a plausible description of the SFH in \jay. Figure \ref{starvol} shows that this scenario requires the assembly of galaxies with stellar masses of $M_{\star}\approx 7\times 10^{11}$~\msun\ by $z \approx 9$. The associated mass of the dark matter (DM) halo is: $M_h \approx (\Omega_m/\Omega_b)\, (M_{\star}+M_g) \approx 4\times 10^{12}$~\msun.  
The comoving number density of such halos derived from the Press-Schechter (PS) approximation \citep{press74} is about $10^{-12}$~\mpc\ \citep{stiavelli09, li07a}. This number is significantly smaller that the $\sim 10^{-9}$~\mpc\ comoving volume density of the $z\sim6$ SDSS~QSOs \citep{fan03} which have optical to near-IR luminosities similar to \jay\ \citep{jiang10}. The apparent three orders of magnitude discrepancy between the observed number density of \jay-like QSOs and that calculated using the PS approximation can probably be attributed to one or more of the following: (1) the breakdown in the accuracy of the formalism at high redshifts and large halo masses (Li07, and references therein); (2) the observations presented by \cite{jiang10} only cover the optical to near-IR wavelengths, which only includes a hot dust component. Their QSO sample may therefore not contain a massive amount of colder dust, and may therefore not be representative of the population of \jay-type objects at far-IR wavelengths; and (3) the Li07 merger scenario is not the only scenario that can lead to the formation of the supermassive black hole, even if the BH-halo mass correlation holds at these redshifts.

In the SN scenario, if we assume that the dust formed in an extremely young burst, the mass of the dark halo associated with an $M_{\star}\approx 10^{11}$~\msun\ galaxy is about $6\times10^{11}$~\msun. The comoving number density of such halos is about $10^{-5}$~\mpc, consistent with the number density of galaxies with similar stellar masses detected in the Great Observatories Origins Deep Survey (GOODS) fields \citep{stark09}. 
If all these galaxies had luminosities similar to \jay, then the comoving bolometric luminosity density at $z\approx 6$ should be about $10^{9}$~\lsun~\mpc. This luminosity density is somewhat larger than the {\it total} luminosity density presented in Figure~18 of \cite{franceschini10}, when extrapolated to $z \approx 6$. However, galaxy count models suggest that galaxies with luminosities above $\sim 2\times10^{13}$~\lsun\ constitute only a small a fraction, about $10^{-4}$, of the total number of galaxies at high redshift \citep{chary01,franceschini10}. This gives a comoving number density of \jay-type objects of $\sim 10^{-9}$~\mpc, similar to the \cite{fan03} observations. So in the SN scenario a significant fraction of the $z\sim 6$ quasars should have massive amounts of dust. This result is not yet supported by current observations, which show that a significant fraction of these objects are dust-free. However, among the $M_{\star}\approx 10^{11}$~\msun\ galaxies, \jay\ is an extremely rare object, consistent with the inferred low extinction and metallicities in $z\sim 7$ galaxies detected in the CDF-South GOODS field with the WFC3/IR camera on the {\it Hubble Space Telescope} \citep{bouwens10}. The rarity of \jay\ in the SN scenario should not be surprising considering the very contrived SFH required to form the inferred mass of dust in this object.

%------------------------------------------------------------
\subsection{The Star Formation Efficiency }
%------------------------------------------------------------

Observations of star forming galaxies revealed an empirical relation between a galaxy's global SFR and its total gas mass.
For a Salpter IMF, this relation can be written as \cite{kennicutt98b}:
\begin{equation}
\label{ }
\Sigma_{SFR}(M_{\odot}\, yr^{-1}\, kpc^{-2}) = (2.5\pm0.7)\times10^{-4}\, \Sigma_g(M_{\odot}\, pc^{-2})^{1.4\pm0.15}
\end{equation}

VLA observations of \jay\ show that the CO gas is distributed over an extended region with an area of $\sim 4$~kpc$^2$ \citep{walter04}. Adopting a total gas mass of $5\times 10^{10}$~\msun, and assuming that it occupies the same area as the CO gas gives an  expected SFR of $\sim 560\pm240$~\myr. For the mass heavy IMF adopted here, the corresponding SFR is $\sim 220\pm90$~\myr. 
This value is larger than the SFR at the epoch of observations in the AGB scenario, suggesting that the SF efficiency at that time is about $\sim 3$ times lower than that in normal star forming galaxies, which varies between 2 and 10\% \citep{kennicutt98b}. 
However, this instantaneous SF efficiency does not reflect its past value which, averaged over a fixed time interval, should scale as $\Sigma_g^{0.4}$. So the average SF efficiency in \jay\ should be similar to that in normal SF galaxies. 

Studies of the relation between the mass  of dark matter (DM) halos and the luminosity of their host galaxies show an evolutionary trend between the galaxies' luminosity function and their halo masses \citep{trenti10}. This trend can be translated into a relation between their SF efficiency, $\eta$, and their halo mass. For masses above $10^{11}$~\msun, the SF efficiency depends only mildly on halo mass, and is given by $\eta \approx 0.06\, (M_h/10^{11}\, M_{\odot})^{0.3}$ \citep{trenti10}. For the inferred halo mass of $4\times 10^{12}$~\msun\ in the AGB scenario, $\eta \approx 0.18$. So the predicted SF efficiency is somewhat higher than the inferred average in the AGB scenario, which can however be attributed to expected large variations in individual objects.

In the SN scenario, all the far-IR luminosity is powered by star formation, suggesting a SFR in excess of $\sim 3000$~\myr. This implies a very high SF efficiency for the given mass of the galaxy. At this rate the galaxy will deplete its reservoir of gas in about 10~Myr, suggesting that we are just witnessing the onset of a large burst of SF in this object, which will be rapidly quenched by stellar feedback.

%====================================
\section{ALTERNATIVE DUST SOURCES}
%====================================
The still unresolved discrepancy between the predicted and inferred stellar masses in the AGB scenario, and the specialized SF scenario required to form the inferred \jay\ dust mass by SNe, raises the need to explore alternative, non-stellar, sources that may produce the large amount of dust in this object. In the following we consider the growth of dust in molecular clouds and their formation around AGN as such sources.

%------------------------------------
\subsection{Molecular Clouds}
%------------------------------------
Numerous observations and theoretical considerations support the notion that dust grains are processed and grow in dense molecular clouds (MCs). Observationally, extinction measurements show that the value of $R_V\equiv A_V/E(B-V)$, an indicator of grain size, increases with hydrogen column density [e.g. \cite{kandori03}, and references therein]. Similarly, interstellar depletions of refractory elements increase with average gas column density \citep{savage96,jenkins09} and infrared spectroscopy of molecular clouds show the presence of ices, expected to have condensed out on interstellar grain cores \citep{van-dishoeck04}. Theoretically, the discrepancy between the relatively short lifetime of interstellar dust \citep{jones04} and the longer lifetime for their replenishment by SNe or AGB stars has been attributed to grain growth in molecular clouds \citep{dwek80b,dwek98, tielens98, zhukovska08}. In particular, the possibility that rapid growth of dust in high-redshift galaxies took place in molecular clouds was considered by \cite{michalowski10a}. Here we provide a more quantitative discussion of such possibility.  
  
Several conditions must be met for grain growth in clouds to be significant: (1) the presence of preexisting refractory grain cores onto which the accretion can take place; (2) the accretion time onto these cores must be shorter than the lifetime of the dense clouds in which this process is taking place.   

Let $\rho_{gr} = n_{gr}\times m_{gr}$ be the mass density of grains in the MC, where $n_{gr}$ is their number density and $m_{gr}$ their mass. Assuming that all grains have the same radius $a$, the growth rate of $\rho_{gr}$ due to accretion of a heavy element $A$ with a molecular weight $\mu_A$ and number density $n_A$ is given by:
\begin{equation}
\label{dmgr_dt}
{d\rho_{gr}\over dt} = \alpha\, \pi a^2\, \mu_A\, n_A\, n_{gr}\, \widetilde{v}
\end{equation}
where $\alpha$ is the sticking coefficient in the collision, and $\widetilde{v} = (8 k T/\pi \mu_A)^{1/2}$ is the mean thermal velocity of $A$ at temperature $T$.  

The accretion time, $\tau_{acc}$, is defined as follows:
\begin{eqnarray}
\label{tau_acc1}
\tau_{acc}^{-1} \equiv {1\over \rho_{gr}}\, {d\rho_{gr}\over dt}  & = & \alpha\, \left({\pi a^2\over m_{gr}}\right)\, \mu_A\, n_A\, \widetilde{v} \\ \nonumber
 & = & \alpha\, \left({3\over 4 \rho a}\right)\, \left({\mu_A\over m_H}\right)\,\left({ n_A\over n_{H_2}}\right)\, m_H\, n_{H_2}\, \widetilde{v} 
\end{eqnarray}

\noindent
Taking the mass of an oxygen atom to represent that of a colliding species, so that $\mu_A/m_H = 16$, and $n_A/n_{H_2} \approx 5\times 10^{-4}$ (assuming solar abundances), we get that:
\begin{equation}
\label{tau_acc}
\tau_{acc}(sec) = {2\times 10^{18}\over \alpha}\, \rho\, \left({a\over \mu m}\right)\, \left({n_{H_2}\over cm^{-3}}\right)^{-1}\, \left({T\over K}\right)^{-1/2} 
\end{equation}
For a grain density of 3~g~cm$^{-3}$, radius of 0.1~\mic, a cloud density of $10^3$~\cc, and cloud temperature of 20~K, the accretion time becomes $\tau_{acc}\sim 3\times 10^6$~yr.

The lifetime, $\tau_c$, of the molecular clouds is given by:
\begin{equation}
\label{tau_cloud}
\tau_c  = {M_c \over \psi}  \approx  {2\times 10^{10}\over 3000} \approx 7\times 10^6\ yr
\end{equation}
where $M_c$ is the MC mass, and where we used the inferred H$_2$ mass and SFR to calculate the average lifetime of the molecular gas.
We used here the large SFR inferred from the SN scenario to minimize the lifetime of the MC. Had we used the SFR from Figure \ref{sfr} the MC lifetime would have been higher by a factor of $\sim 50$. Within the uncertainties of the chosen values of the various parameters, the cloud lifetime is not significantly shorter than the accretion time. Consequently grain growth in the dense ISM can play an important role in determining the observed dust mass.

The accreted material may be in the form of organic refractory material \citep{greenberg95}. The main observational constraints, extinction, diffuse IR emission, interstellar abundances, in the Milky Way can also be met with composite dust models that incorporate such material \citep{li97,zubko04}. If so, then the dust of \jay\ could also consist of such composite grains that could have a significant abundance in the diffuse ISM. In the Milky Way, most (60--75\%) of the IR emission is radiated by dust in the diffuse ISM, with only 15--30\% radiated from MCs \citep{sodroski97}. 
Likewise, the IR emission from \jay\ could arise either from the dust in the MCs, or from dust in the diffuse ISM, including H~II regions. The origin of the IR emission in this case can therefore only be determined by considering dust evolution models that incorporate a two phase medium, and take the physical processes and the cycling between the phases into account. 

%------------------------------------
\subsection{``Smoking Quasars"} 
%------------------------------------
The concept of ``smoking quasars'', that is, the possibility that quasars may be producers of interstellar dust was first put forward by \cite{elvis02}. In his scenario, the broad emission line clouds (BELCs), confined by the  centrifugally driven outflowing wind \citep{kartje96}, go through the same density and temperature conditions that characterize the dusty winds in stellar outflows. It is therefore  naturally to postulate that if the conditions in the stellar winds are ripe to form dust, a similar process will be taking place in the BELCs.
This idea was further discussed by \cite{maiolino06}.

The main issue, not addressed in these papers is whether the BELCs were dusty to start with \citep{elitzur06}. If so, then the IR emission observed in the vicinity of the AGN is not the signature of radiation from newly-formed dust, but that of reprocessed dust. This dust could have formed previously in SNe or dusty stellar winds. If so, quasars cannot be considered as net producers of interstellar dust. The issue of quasars as dust sources can therefore only be settled once we have a clearer understanding the origin of the BELCs, and of the dust formation process in these objects. 
 
%==================================
\section{DISCUSSION AND SUMMARY}
%==================================
Using our new integral solutions for the chemical evolution of the ISM gas and its metal and dust contents, we examined the origin of the large amount of dust discovered in the high$-z$ quasar \jay. 
We have confirmed that AGB stars can produce the required amount of dust in a merger scenario in which the assembled galaxy undergoes a period of intense star formation at $z \approx 9$, followed by a period of significantly lower stellar activity. In this AGB scenario, an average 20~\msun\ supernova needs to make at least $\sim 1$~\msun\ of dust in order to be a viable source of dust at this high redshift. However, SNe can yet produce the required amount of dust with a significantly lower dust yield, provided the galaxy underwent a very special SFH. In the following we summarize the main results of our paper, and discuss future observations and studies needed to distinguish between the two scenarios in order to determine the origin and nature of the dust in \jay\ and similar objects in the early universe.
\begin{enumerate}
\item {\bf The AGB scenario} -  A galaxy must be at least 400~Myr old for AGB stars to have made a significant contribution to its reservoir of dust, a conditioned fulfilled by the merger scenario for the formation of \jay. However, this SFH is not unique, and we outlined several alternative scenarios for the formation of large amounts of dust with AGB stars. At any given epoch, the intensity of the starburst must be at least 2000~\myr\ in order to produce the inferred amount of dust, and the star formation rate must have dropped to less than $\sim 50$~\myr\ thereafter so that the dust that is injected at the epoch of observations will not be significantly destroyed. 
\item {\bf The SN scenario} - In general, an average SN must condense about 1~\msun\ of dust in order to produce the observed amount of dust in \jay. Our current state of knowledge, observationally and theoretically, suggest that the explosive ejecta of a progenitor $\sim 20$~\msun\ star produces only about $0.1-0.15$~\msun\ of dust. Such progenitors are the most common metal producing SNe, and fall short of meeting the production requirement. However, our knowledge is based on limited number of observations and calculations, so it may change in the future. Similarly to the alternative AGB scenario, SNe can produce the observed amount of dust with a dust yield of $\sim 0.15$~\msun\ in a single short-duration and intense burst of star formation, followed by a period of very low star forming activity. Such scenarios must be very rare in order to explain the low comoving number density of hyperluminous IR objects at high redshifts.
\item {\bf The stellar mass and gas content of \jay} - The stellar mass offers an important integral constraint on the duration and intensity of the burst of star formation. These place integral constraints on the total amount of dust that could have formed in the quasar. In the SN scenario $M_{\star}\approx 10^{11}$~\msun\ of stars are formed around $z\approx 7$.  So if dust formation was very efficient in these objects, the SN scenario would  produce more dusty hyperluminous IR stars than currently observed at high redshifts. The formation of SN dust must therefore be a very rare event in these objects, limited by the dust formation and grain destruction efficiencies in these objects. The AGB scenario requires the formation of $M_{\star}\approx 7\times10^{11}$~\msun\ at redshifts of $\sim 8-9$, suggesting that \jay\ is a very rare event, but perhaps consistent with the number of \jay-like QSOs detected at $z\sim 6$. If so, a significant fraction of $z\sim 6$ quasars should have massive amount of dust, a prediction that is yet unsupported by current observations. Furthermore, the stellar mass produced in the AGB scenario is much larger that the inferred dynamical mass of the galaxy. Resolving this discrepancy is a crucial step in determining the origin of the dust in \jay. 
\item{\bf The nature of the optical to near-IR SED of \jay} - In the SN scenario all the emission in this wavelength region is produced by stars, whereas in the AGB scenario all this emission is produced by the AGN. In the absence of a detailed spectrum, we used the slope of the SED to discriminate between the two scenarios. The results show that the slope of the SED is typical of quasars, and also consistent with that produced by an extincted burst of star formation. The detection of mid-IR fine structure lines that serve as unambiguous diagnostics of AGN or starburst activity will be extremely useful for distinguishing between the two scenarios. 
\item {\bf The efficiency of grain destruction in the ISM} - Our current state of knowledge is predominantly based on the simulations that were performed for a homogenous ISM. These calculations need to be expanded for a 2-phase ISM, taking also the spatial distribution of dust sources and sinks into account. For example, one can easily envision an intense starburst or spatially isolated SNe producing large amount of dust in one location in the galaxy, followed sometime later by another physically isolated starburst that will have little effect on the destruction of the dust created in the first. Spatially isolated starburst or SNe may therefore increase the currently estimated short lifetime of the dust in the ISM.
\item {\bf Dust growth in the ISM} - Accretion in molecular clouds must play an important role in explaining the origin of the Galactic pattern of interstellar depletions. A necessary condition for this process to be important is that the lifetime of molecular clouds be larger than the timescale for grain growth by accretion. This condition is met in \jay, rendering it a possible, but not necessary, source for the growth of dust in this object. Detailed information of the correlation of the IR emission with the different gas phases in \jay\ may resolve this ambiguity.
\item {\bf Dust formation in around AGN} - The effectiveness of AGNs as sources of newly-formed dust remains unclear. The origin of the dense broad-line-emitting clouds, and whether they were initially dusty so that the AGN merely reformed pre-existing dust, are questions that need to be addressed to ascertain their role as dust sources in quasars.     
\end{enumerate}
{\bf Acknowledgements}
ED  thanks the Department of Astronomy at the University of Maryland, and the Department of Astronomy and Astrophysics at Tel Aviv University for their hospitality during the time of the writing of this paper. ED also acknowledges helpful discussions with Richard Mushotzky, Sylvain Veilleux,  Amiel Sternberg, Dan Laor, and Hagai Netzer, and thanks Rick Arendt and Dieter Hartmann for comments on parts of the manuscript. ED also thanks Massimo Stiavelli for providing an expanded version of Figure 2.6 in his book for use in this study. Finally, we thank the anonymous referee for his/her useful comments that have led to significant improvements in the paper.

\newpage
% \bibliographystyle{/Users/edwek/Library/texmf/tex/latex/misc/aastex52/aas.bst}
% \bibliographystyle{ /pub/softw/lib/tex.local/aastex52/aas.bst}
%\bibliography{/Users/edwek/science/Bib_Desk/Astro_BIB.bib}
 \bibliographystyle{/pub/softw/lib/tex.local/aastex52/aas.bst}
% \bibliography{/Volumes/Apps_and_Docs/edwek/science/00-Bib_Desk/Astro_BIB.bib}
 \bibliography{/Users/edwek/science/00-Bib_Desk/Astro_BIB.bib}
% \bibstyle{ /pub/softw/lib/tex.local/aastex52/aas.bst}
%\bibdata{/Users/edwek/science/Bib_Desk/Astro_BIB.bib}

\newpage

%%==== table 1
%\begin{deluxetable}{lcc}
%\tablewidth{0pt}
%\tablecaption{Select Properties of QSO J1148+5152}
%\tablehead{
%\colhead { Observed quantity} &
%\colhead { Value} &
%\colhead {Reference}   
%}
%\startdata
%% \begin{table}[h]
%%  \centering
%%  \begin{tabularx}{\textwidth}{XXr}
%%    \hline\hline
%      $z$   & $6.4 \pm0.03$            & 1         \\
%	$L_{bol} $~(\lsun)      & $1.4\times 10^{14}$   & 1\\
%	$L_{IR}$(\lsun) 	& $2\times 10^{13}$ & 1 \\
%%    \hline
%     $M_{gas}$(\msun) & $\sim2\times10^{10}$   & 1   \\
%      $M_{dust}$(\msun)  & $\sim 3\times10^8$    & 1 \\
%      $M_{dyn}$(\msun)   & $(5.0\pm2.5)\times10^{10}$  & 1
%%       & & \\
%\enddata
%\tablenotetext{1}{ \citep{white03,iwamuro04,bertoldi03b}}
%\tablenotetext{1}{ \citep{dwek07b,beelen06}}
%\tablenotetext{1}{ \citep{walter04,bertoldi03b}}
%\tablenotetext{1}{ \citep{walter04}}
%\tablenotetext{1}{ \citep{walter04}}
%\label{j11_obs}
%\end{deluxetable}

%==== table 1
\begin{deluxetable}{cccccc}
\tablewidth{0pt}
\tabletypesize{\footnotesize}
\tablecaption{Main-sequence Lifetimes of AGB Progenitor Stars\tablenotemark{1}}
\tablehead{
\colhead{Mass } &
\multicolumn{5}{c}{   Metallicity   } \\
%\hline
\cline{2-6}
\colhead {(\msun)} &
\colhead {0.0001} &
\colhead {0.001} &
\colhead {0.008} &
\colhead {0.0100} &
\colhead {0.02}  
}
\startdata
1.0& 4998 & 6774 & 8634 & 8846 & 9516. \\
1.5& 1474 & 2022 & 2415 & 2446 & 2528. \\
2.0&  663 &  914 & 1047 & 1054 & 1062. \\
2.25&  486 &  670 &  756 &  759 &  757. \\
2.5&  371 &  512 &  570 &  570 &  565. \\
2.75&  292 &  404 &  444 &  443 &  436. \\
3.0&  236 &  326 &  355 &  354 &  346. \\
3.25&  195 &  270 &  291 &  290 &  282. \\
3.5&  164 &  227 &  243 &  241 &  234. \\
3.75&  140 &  194 &  206 &  204 &  197. \\
4.0&  121 &  168 &  177 &  175 &  168. \\
4.5&   94 &  129 &  135 &  134 &  128. \\
5.0&   75 &  104 &  107 &  106 &  100. \\
5.5&   62 &   85 &   87 &   86 &   81. \\
6.0&   52 &   72 &   73 &   72 &   68. \\
6.5&   45 &   61 &   62 &   61 &   57. \\
7.0&   39 &   53 &   53 &   53 &   49. \\
7.5&   34 &   47 &   47 &   46 &   43. \\
8.0&   31 &   42 &   41 &   41 &   38. \\
\enddata
\tablenotetext{1}{~Lifetimes in Myr, calculated using the analytical formulae from \cite{raiteri96}.}
\label{ms_lifetimes}
\end{deluxetable}

%===== Table 3 ========
\begin{deluxetable}{ll}
\tablewidth{0pt}
\tabletypesize{\footnotesize}
\tablecaption{Energy Output from \jay}
\tablehead{
\colhead {Wavelengths} &
\colhead {Luminosity (\lsun)} 
}
\startdata
 UV--near-IR	(0.1--5 \mic)& $(5.7 \pm1.3)\times10^{13}$ \\ 
 mid-IR (5--30 \mic)	& $(1.8 \pm0.2)\times10^{13}$ \\ 
 far-IR	 (30--300 \mic) & $(2.5 \pm0.5)\times10^{13}$ \\ 
bol (0.1--300 \mic)	& $(10.0 \pm1.5)\times10^{13}$ \\ 
\hline
calculated stellar\tablenotemark{1} &  $1.6\times 10^{13}$ \\
\enddata
\tablenotetext{1}{The total bolometric output at $t=900$~Myr (see Figure \ref{starvol}).}
\label{lum}
\end{deluxetable}

\end{document}